\documentclass[twocolumn,english,aps,prd,reprint,floatfix,notitlepage,nofootinbib,preprintnumbers,superscriptaddress]{revtex4-1}
\pdfoutput=1
\usepackage{lmodern}

\usepackage[T1]{fontenc}
\usepackage[latin9]{inputenc}
\usepackage{geometry}
\usepackage{subfigure,lmodern, amsmath,amssymb, graphicx, pifont, adjustbox, bm, xcolor}
\usepackage{amsfonts}
\usepackage{geometry}
\usepackage{comment}
\usepackage{mathtools}
\usepackage{float}
\usepackage{slashed}
\usepackage{ragged2e}
\usepackage{tikz}
\usepackage{array}
\usepackage{hhline}
\usepackage{dsfont}

\makeatletter\g@addto@macro\bfseries{\boldmath}\makeatother

\usepackage{stackengine}
\usepackage{esint}
\usepackage[unicode=true,pdfusetitle,
 bookmarks=true,bookmarksnumbered=false,bookmarksopen=false,
 breaklinks=false,pdfborder={0 0 1},backref=false,colorlinks=true]
 {hyperref}
\hypersetup{
 pdfauthor={Clifford Cheung, Grant N. Remmen},
 citecolor=black,linkcolor=black,urlcolor=black}

\newcommand{\appendixref}[1]{\hyperref[#1]{appendix~\ref{#1}}}
\def\equationautorefname~#1\null{eq.\,(#1)\null}
\usepackage{breakurl}
\usepackage{breakurl}
\usepackage[hang,flushmargin]{footmisc} 
\allowdisplaybreaks
\makeatletter

\usepackage{etoolbox}
\apptocmd{\thebibliography}{\justifying\setlength{\leftskip}{7.4mm}}{}{} 
 
 \usepackage{relsize}
\usepackage{babel}

\makeatletter
\def\simgt{\mathrel{\lower2.5pt\vbox{\lineskip=0pt\baselineskip=0pt
           \hbox{$>$}\hbox{$\sim$}}}}
\def\simlt{\mathrel{\lower2.5pt\vbox{\lineskip=0pt\baselineskip=0pt
           \hbox{$<$}\hbox{$\sim$}}}}
\makeatother

\usepackage{changepage}

\makeatletter
\def\l@subsubsection#1#2{}
\makeatother

\newcommand{\be}{\begin{equation}}
\newcommand{\ee}{\end{equation}}
\newcommand{\bea}{\begin{eqnarray}}
\newcommand{\eea}{\end{eqnarray}}
\newcommand{\Fig}[1]{Fig.~\ref{#1}}

\newcommand{\Eq}[1]{Eq.~(\ref{#1})}
\newcommand{\Eqs}[2]{Eqs.~(\ref{#1}) and (\ref{#2})}
\newcommand{\Sec}[1]{Sec.~\ref{#1}}

\newcommand{\eq}[2]{\be\begin{aligned}#1 \label{#2}\end{aligned}\ee}
\newcolumntype{P}[1]{>{\centering\arraybackslash}p{#1}}

	\definecolor{dartmouthgreen}{rgb}{0.05, 0.5, 0.06}

\begin{document}

\newcommand{\SixPtLadder}{
\begin{tikzpicture}[baseline=-0.1cm, line width=0.8pt, scale=0.25]

  % Define horizontal points
  \coordinate (A) at (0,0);
  \coordinate (B) at (3,0);
  \coordinate (C) at (6,0);
  \coordinate (D) at (9,0);

  % Draw horizontal backbone with labels below
  \draw (A) -- node[below, yshift=-6pt] {$n$} (B);
  \draw (B) -- node[below, yshift=-6pt] {$n$} (C);
  \draw (C) -- node[below, yshift=-6pt] {$n$} (D);

  % Vertical upward legs (3 and 4)
  \draw (3,0) -- ++(0,2.4);  % leg 3
  \draw (6,0) -- ++(0,2.4);  % leg 4

  % Diagonal legs on the left
  \draw (A) -- ++(-2,2);   % leg 1
  \draw (A) -- ++(-2,-2);  % leg 2

  % Diagonal legs on the right
  \draw (D) -- ++(2,2);    % leg 5
  \draw (D) -- ++(2,-2);   % leg 6

\end{tikzpicture}
}

\newcommand{\FivePtLadder}{
\begin{tikzpicture}[baseline=-0.1cm, line width=0.8pt, scale=0.25]

  % Define horizontal points for a 2-segment ladder
  \coordinate (A) at (0,0);
  \coordinate (B) at (3,0);
  \coordinate (C) at (6,0);

  % Draw horizontal spine with "n" labels below
  \draw (A) -- node[below, yshift=-6pt] {$n$} (B);
  \draw (B) -- node[below, yshift=-6pt] {$n$} (C);

  % Vertical upward leg (leg 3)
  \draw (3,0) -- ++(0,2.4);

  % Diagonal legs on the left (legs 1 and 2)
  \draw (A) -- ++(-2,2);   % leg 1
  \draw (A) -- ++(-2,-2);  % leg 2

  % Diagonal legs on the right (legs 4 and 5)
  \draw (C) -- ++(2,2);    % leg 4
  \draw (C) -- ++(2,-2);   % leg 5

\end{tikzpicture}
}

\newcommand{\FourPtLadder}{
\begin{tikzpicture}[baseline=-0.1cm, line width=0.8pt, scale=0.25]

  % Define horizontal points for a 1-segment ladder
  \coordinate (A) at (0,0);
  \coordinate (B) at (3,0);

  % Draw horizontal spine with "n" label below
  \draw (A) -- node[below, yshift=-6pt] {$n$} (B);

  % Diagonal legs on the left (legs 1 and 2)
  \draw (A) -- ++(-2,2);   % leg 1
  \draw (A) -- ++(-2,-2);  % leg 2

  % Diagonal legs on the right (legs 3 and 4)
  \draw (B) -- ++(2,2);    % leg 3
  \draw (B) -- ++(2,-2);   % leg 4

\end{tikzpicture}
}

\preprint{CALT-TH 2025-010}

\title{Multipositivity Bounds for Scattering Amplitudes}

\author{Clifford Cheung}
\affiliation{Walter Burke Institute for Theoretical Physics, California Institute of Technology, Pasadena, CA 91125}
\author{Grant N.~Remmen}
\affiliation{Center for Cosmology and Particle Physics, Department of Physics, New York University, New York, NY 10003}    
    
\begin{abstract}

\noindent Lorentz invariance, unitarity, and causality enforce powerful constraints on the theory space of physical scattering amplitudes.  However, virtually all efforts in this direction have centered on the very simplest case of four-point scattering.  In this work, we derive an infinite web of ``multipositivity bounds'' that nonlinearly constrain all tree-level higher-point scattering amplitudes under similarly minimal assumptions.   Our construction rules out several deformations of the string and implies mixed-multiplicity bounds on  the Wilson coefficients of planar effective field theories.   Curiously, an infinite class of multipositivity bounds is exactly saturated by the amplitudes of the open string.
\vspace*{7mm}
\end{abstract}
\maketitle

\setcounter{secnumdepth}{2}
\setcounter{tocdepth}{2}
\tableofcontents

\vspace*{3mm}

\section{Introduction}

The laws of physics are far from arbitrary: not every Lagrangian defines a universe of self-consistent dynamics, and not every function of momentum defines the scattering amplitude of a physically realizable process.  In fact, fundamental limits on theory space can be rigorously derived from bedrock principles like Lorentz invariance, unitarity of quantum mechanics, and causality of signal propagation \cite{Adams:2006sv, Pham:1985cr,Ananthanarayan:1994hf,Pennington:1994kc,Jenkins:2006ia,Dvali:2012zc}. 
The amplitudes bootstrap program aims to demarcate the boundary between consistent and inconsistent effective field theories (EFTs) implied by these modest assumptions.  

In recent years, this effort has welcomed numerous developments, including the EFT-hedron~\cite{Arkani-Hamed:2020blm}, null constraints \cite{Caron-Huot:2020cmc, Caron-Huot:2021rmr,Tolley:2020gtv}, and moment bounds \cite{Bellazzini:2020cot}. These insights have revealed a complex network of constraints on the Wilson coefficients of EFTs.  Broadly referred to as {\it positivity bounds}, these constraints have found many applications, including to gravity~\cite{Bellazzini:2015cra,Cheung:2016wjt,Camanho:2014apa,Gruzinov:2006ie,Arkani-Hamed:2021ajd,Cheung:2014ega,Bellazzini:2019xts,Andriolo:2020lul,Caron-Huot:2021rmr,Caron-Huot:2022ugt,Caron-Huot:2022jli,Cheung:2016yqr,deRham:2017xox,Camanho:2016opx,Bellazzini:2020cot,Bern:2021ppb,Tolley:2020gtv}, string theory~\cite{Berman:2023jys, Berman:2024wyt, Albert:2024yap, Berman:2024eid,Cheung:2022mkw,Cheung:2023adk,Cheung:2023uwn,Cheung:2024uhn,Cheung:2024obl,Haring:2023zwu,Arkani-Hamed:2022gsa,Bhardwaj:2024klc}, cosmology~\cite{Freytsis:2022aho,Green:2023ids,Baumann:2019ghk}, and  phenomenology~\cite{Remmen:2019cyz,Remmen:2020vts,Remmen:2020uze,Remmen:2022orj,Remmen:2024hry,Bi:2019phv,Zhang:2018shp,Low:2009di,Englert:2019zmt,ZZ,YZZ,Trott,HYZZ,Chala:2023xjy}.

Despite immense progress, the lion's share of work on this subject suffers from the same persistent drawback: a restriction to scattering amplitudes with {\it exactly four external legs}.  At the same time, it strains credulity to imagine that the full span of constraints on theory space arises solely from four-point analyses.   
Higher-point scattering constraints nonetheless remain essentially unexplored, modulo efforts to derive bounds from factorization \cite{Arkani-Hamed:2023jwn}, which requires explicit specification of the spins and degeneracies in the spectrum, or from a version of the higher-point optical theorem, which requires strong assumptions about lower-point scattering~\cite{Chandrasekaran:2018qmx}. 

In this work, we present a new framework for deriving positivity bounds at arbitrary multiplicity, which we dub {\it multipositivity bounds}.  These constraints are mixed-multiplicity inequalities that nonlinearly constrain the residues of amplitudes and the Wilson coefficients of EFTs.
Notably, our results make no assumptions about the spectrum of masses and spins, in particular with respect to the presence or absence of degeneracies.

The underlying logic of our multipositivity bounds is as follows.  In the forward limit, the residue of the tree-level four-point amplitude on an exchanged resonance is
\eq{
R_n \;\; =\;\; \FourPtLadder \;\; =\;\;  \langle n| n\rangle,
}{R4diagram}
where the level $n$ labels the set of states in the spectrum with mass $m_n$.
As we will show, this residue is mathematically equal to the expectation value of the identity on an unnormalized vector $|n\rangle$ describing the intermediate state produced by the pair of external particles.  The residues at level $n$ of the five-point amplitude,
\eq{
R_{nn} \;\; =\;\; \FivePtLadder \;\;=\;\;  \langle n| {\cal O}_n |n\rangle,
}{R5diagram}
and of  the six-point amplitude,
\eq{
R_{nnn} \;\; =\;\; \SixPtLadder \;\;  =\;\;  \langle n| {\cal O}_n^2 |n\rangle,
}{R6diagram}
are similarly equal to inner products.  Here  ${\cal O}_n$ is an operator parameterizing the unknown three-point amplitudes inserted along the spines of these half-ladder topologies.  Since all couplings are real numbers, we will find that ${\cal O}_n$ is Hermitian.
Consistency of the inner product then implies that $\langle n| n\rangle\geq 0$ and $
 \langle n| {\cal O}_n^2 |n\rangle\geq 0$, so
\eq{
R_n\geq 0 \;\; \textrm{and}\;\; R_{nnn} \geq 0.
}{R_diag_bound}
Moreover, the variance $ \langle n|  {\cal O}_n^2 |n \rangle \langle n|n\rangle - \langle n|  {\cal O}_n |n \rangle^2 \geq 0$ is positive semidefinite, which implies the inequality 
\begin{widetext}
\eq{
\left( \SixPtLadder\right) \left(\FourPtLadder\right) -\left(\FivePtLadder\right)^2 =R_{nnn} R_n - R_{nn}^2  \geq 0.
}{R_det_bound}
\end{widetext}
The physical origin of these constraints is quite intuitive.  By restricting all intermediate legs to level $n$, the scalars inserted along the half-ladder carry zero momentum and are thus soft.  Consequently, $R_n$, $R_{nn}$, and $R_{nnn}$ describe the propagation of a level $n$ state in a soft background.  Since ${\cal O}_n$ defines a soft correction to the propagator of the level $n$ state, it is automatically Hermitian.  

As we will see, the inequalities in \Eqs{R_diag_bound}{R_det_bound} are the very simplest representatives of an infinite tower of multipositivity bounds that extend to arbitrarily high multiplicity, across different levels of the spectrum, and even away from the forward limit.

Our multipositivity bounds on residues imply new constraints on the low-energy Wilson coefficients of a scalar EFT that arises from a tree-level ultraviolet completion.\footnote{Similarly, the EFT-hedron bounds are also derived in a tree-level context, though as noted in Ref.~\cite{Arkani-Hamed:2020blm} a four-point loop-level completion can always be written in terms of integrals over trees. If such a construction extends to $N$-point scattering, then our EFT bounds apply to any weakly coupled completion.}  In particular, we expand the EFT amplitudes in powers of energy, focusing on the Wilson coefficients of particular kinematic structures, 
\eq{
A_4 &=  \sum_k g_{4,k} s_{1} ^k +\cdots \\
A_5 &=  \sum_k g_{5,k} s_{1}^k s_{2}^k +\cdots \\
A_6 &=  \sum_k g_{6,k} s_{1}^k s_{2}^k s_{3}^k +\cdots ,
}{}
where $s_i$ are the kinematic invariants entering into the propagators along the spine of each half-ladder topology.  
Our multipositivity bounds imply constraints such as
\eq{
g_{6,k} g_{4,k} - g_{5,k}^2 \geq 0,
}{}
along with their higher-point generalizations.

As a consistency check of our multipositivity bounds on residues, we construct {\it all possible} half-ladder residues consistent with exchanges of any spin mediated by arbitrary three-point interactions and verify the validity of our bounds up to multiplicity $N\,\,{=}\,\,8$ for spin ${\leq}\,\, 2$ and $N\,\,{=}\,\,6$ for spin ${\leq}\,\, 3$.  We then apply our bounds to constrain---and in some cases rule out---various proposed deformations of the $N$-point amplitudes of open string theory, including certain ``bespoke'' dual resonant amplitudes with tunable spectra~\cite{Cheung:2023uwn} and a class of worldsheet deformations of the string~\cite{Gross:1969db}.

\section{Positivity of Residues}

\subsection{Four-Point Scattering}

The four-point color-ordered amplitude $A(s,t)$ is a function of the kinematic invariants
\eq{
s= -(p_1+p_2)^2 \quad \textrm{and} \quad t = -(p_2+p_3)^2,
}{}
working throughout in mostly-plus metric signature and cyclic ordering of particle labels.
The residue of $A(s,t)$ on a resonance is
\eq{
R_n(t) = \lim_{s\rightarrow m_n^2} (m_n^2-s) A(s,t),
}{res_def}
where $m_n^2$ is the mass squared of the level $n$ state. We assume throughout that the $n=0$ level is massless,
\eq{
m_0^2=0,
}{}
and that this level includes the scalar that defines our external states.  We take the spectrum to be tachyon-free, $m_n^2\geq 0$, but otherwise arbitrary. 

Next, let us define a set of basis vectors $|n,\ell\rangle$ spanning the level $n$ subspace.  Here $\ell$ labels all degenerate states at level $n$. For example, $\ell$ implicitly encodes all spin, charge, or flavor quantum numbers.  Without loss of generality, we take these states to be orthonormal,
\eq{
  \langle n,\ell |n',\ell'\rangle = \delta_{nn'} \delta_{\ell \ell'}.
  }{}
Note that $|n,\ell\rangle$ is {\it not} a Fock state as conventionally defined in quantum field theory.  Indeed, a genuine Fock state exhibits delta function normalization and carries a momentum vector label, which is not the case here.  Instead, we take $|n,\ell\rangle$ to define a state at level $n$ at some {\it fixed} reference momentum $q_n$.  For instance, $q_n$ could be at rest if the particle is massive or pointing in some chosen direction if it is massless.  The precise choice of reference $q_n$ is unimportant, and we have introduced it only to ensure that $|n,\ell\rangle$ describes a unique state.

The tree-level manifestation of unitarity is {\it factorization}, which says that the residue in \Eq{res_def} is equal to\footnote{
We work in a convention in which the color-ordered amplitude corresponds to the color factor $-f^{abe} f^{cde}$.  The minus sign is needed so that in the forward limit, for which $a=d$ and $b=c$, the color factor $-f^{abe}f^{bae} = f^{abe} f^{abe} >0$ is positive.}
\eq{
R_n(t) = -\sum_\ell  A_{n}^{\ell}(p_1,p_2)A_{n}^{\ell}(p_3,p_4),
}{R_AA}
where $A_{n}^{\ell}(p_1,p_2)$ is the color-ordered three-point amplitude\footnote{Here the first and second arguments of this three-point amplitude are both defined as in-states.} describing the fusion of a pair of massless scalars of momenta $p_1$ and $p_2$ into the state $|n,\ell\rangle$.  Under exchange of the scalars,  $A_{n}^{\ell}(p_1,p_2) = -A_{n}^{\ell}(p_2,p_1)$ flips under an overall sign, while under complex conjugation, $\bar A_{n}^{\ell}(p_1,p_2)  = A_{n}^{\ell}(-p_1,-p_2) $ flips the signs of its momentum arguments.  
Unitarity requires reality of all three-point couplings, in which case Eq.~\eqref{R_AA} is equivalent to the statement of nonnegativity of the partial wave expansion of the four-point residue.

\medskip

\noindent {\it   Forward Kinematics.} Now let us first consider the forward limit for real kinematics, where
\eq{
p_3 = - p_2 \quad \textrm{and} \quad p_4 = -p_1,
}{}
corresponding to $t=0$.   Starting from \Eq{R_AA}, we then use that $A_{n}^{\ell}(p_3,p_4)= A_{n}^{\ell}(-p_2,-p_1)=-\bar A_{n}^{\ell}(p_1,p_2)$ to express the residue as a sum of squares, 
\eq{
R_n(0) = \sum_\ell  |A_{n}^{\ell}(p_1,p_2)|^2.
}{}
At this point it will be convenient to define a vector describing the superposition of level $n$ states  that are produced by the merging scalars,
\eq{
 | n \rangle &=  \sum_\ell  A_{n}^{\ell}(p_1,p_2) |n,\ell\rangle,
}{n_ket}
the Hermitian conjugate of which is
\eq{
 \langle n | &=  \sum_\ell  \bar A_{n}^{\ell}(p_1,p_2) \langle n,\ell |.
}{}
Unlike $| n,\ell\rangle$, the state $|n\rangle$ is {\it not} unit-normalized, and furthermore its structure is dictated entirely by the dynamics of the three-point amplitude.  In terms of this state, the residue in the forward limit is
\eq{
 R_n(0)=\langle n| n\rangle   \geq 0,
}{}
which is manifestly nonnegative.

\medskip

\noindent {\it  Complex Kinematics.}  Stronger positivity bounds can be derived using a certain complexification of the forward limit.  In particular, consider a shift of the momentum reminiscent of on-shell recursion relations~\cite{BCFW},
\eq{
p_1(z)&=p_1+zv, \qquad &p_2(z)&=p_2-zv, \\
p_{3}(\bar z) &= -p_2+\bar z\bar v, & p_4(\bar z) &= -p_1 -\bar z\bar v.
}{BCFW}
This shift defines a ``complex forward limit'' for which crossing acts on momenta as a sign flip together with complex conjugation.  Here
 $z$ is a complex number, and $v$ is a vector satisfying
\eq{
p_i v = p_i \bar v =v^2 = \bar v^2 = 0,
}{qq}
with normalization
\eq{
\quad v\bar v = \frac{1}{2}.
}{}
For real $p_1$ and $p_2$, the null vector $v$ is complex and spacelike. Note that this shift induces a momentum transfer,  
 \eq{
t(z,\bar z)&=  -( p_2(z)+p_3(\bar z))^2 = z\bar z  \geq 0,
}{t_from_zzbar}
in which case the residue becomes
\eq{
R_n(t(z,\bar z)) = \sum_\ell  |A_{n}^{\ell}(p_1(z), p_2(z))|^2.
}{}
Crucially, the above expression is {\it still} a sum of squares. It is ``forward'' in the sense that the initial and final states are complex conjugates of each other, but nevertheless manages to probe $t> 0$. To see why this is, let us define a generalization of the $|n\rangle$ vector in Eq.~\eqref{n_ket},
\eq{
 | n,z\rangle &=  \sum_\ell  A_{n}^{\ell}(p_1(z), p_2(z)) |n,\ell\rangle ,
}{nz_ket}
whose Hermitian conjugate is\footnote{Note that $|n,z\rangle=(\langle n,\bar z|)^\dagger$ are Hermitian conjugates, so they correspond to the same value of $z$.  In an abuse of notation we label  $\langle n,\bar z|$ by $\bar z$ to make the dependence on $\bar z$ manifest.}
\eq{
 \langle n, \bar z | &=  \sum_\ell  \bar A_{n}^{\ell}(p_1(\bar z), p_2(\bar z)) \langle n,\ell | .
}{nzbar_ket}
Even though $ | n,z\rangle$ is not unit-normalized, we emphasize that it is a {\it physical state}
because it is a linear combination of the physical states $|n,\ell\rangle$.  In particular, the complexification of $p_1(z)$ and $p_2(z)$ has simply modified the wave function coefficients of each $|n,\ell\rangle$ branch.  In terms of these vectors, the residue is
\eq{
R_n(t(z,\bar z)) = \langle n, \bar z| n,z\rangle   \geq 0, 
}{}
which directly implies that
\eq{
R_n(t)  \geq 0\quad \textrm{for} \quad t\geq 0.
}{}
We can also construct new physical states from $|n,z\rangle$ by taking derivatives with respect to $z$, 
\eq{
\partial_z^r | n,z\rangle  &=  \sum_\ell \partial_z^rA_{n}^{\ell}(p_1(z), p_2(z)) |n,\ell\rangle,
}{}
with the analogous Hermitian conjugate,
\eq{
\partial_{\bar z}^r\langle n, \bar z | &=  \sum_\ell  \partial_{\bar z}^r  \bar A_{n}^{\ell}(p_1(\bar z), p_2(\bar z)) \langle n,\ell |.
}{}
From \Eq{t_from_zzbar} we observe the general identity,
\eq{
\partial_{ z}^r\partial_{\bar z}^r (\cdots)|_{z=\bar z=0} = r!\,\partial_t^r (\cdots)|_{t=0}.
}{}
Hence, the $r$th derivative of $R_n(t)$ with respect to $t$, evaluated at $t=0$, is
\eq{
R_n^{(r)}(0) = \frac{1}{r!} \left( \partial_{\bar z}^r \langle n, \bar z|  \right)\left( \partial_{z}^r | n,z\rangle \right)|_{z=\bar z =0} \geq 0 ,
}{R_nr_pos}
which is also manifestly nonnegative, and which in turn implies that $R_n(t)$ is a monotonic function of $t\geq 0$.  This statement is of course well known on account of the mathematical properties of the Gegenbauer polynomials in the partial wave expansion of the scattering, but the above story provides physical insight into this fact.

\begin{figure*}[t]
\centering
\begin{tikzpicture}[line width=0.8pt, scale=0.6]

  % Coordinates for horizontal spine
  \coordinate (A) at (0,0);
  \coordinate (B) at (3,0);
  \coordinate (C) at (6,0);
  \coordinate (D) at (9,0);
  \coordinate (E) at (12,0);
  \coordinate (F) at (15,0);

  % Horizontal spine with medium-size labels
  \draw (A) -- node[below, yshift=-6pt] {\scalebox{1.2}{$\textstyle n_1$}} (B);
  \draw (B) -- node[below, yshift=-6pt] {\scalebox{1.2}{$\textstyle n_2$}} (C);
  \draw (C) -- (D);
  \draw (D) -- (E);
  \draw (E) -- node[below, yshift=-6pt] {\scalebox{1.2}{$\textstyle n_{N-3}$}} (F);

  % Aligned ellipses
  \node at (9,1.4) {\scalebox{1.2}{$\cdots$}};   % upper
  \node at (9,-1.4) {\scalebox{1.2}{$\cdots$}};  % lower

  % Vertical legs
  \draw (3,0) -- ++(0,2.4);   % at B
  \draw (6,0) -- ++(0,2.4);   % at C
  \draw (12,0) -- ++(0,2.4);  % at E

  % Diagonal legs (true length = 2.4), with swapped labels
  \draw (A) -- ++(-1.697,-1.697) node[left, xshift=-4pt] {\scalebox{1.2}{$p_1$}}; % bottom left leg
  \draw (A) -- ++(-1.697, 1.697) node[left, xshift=-4pt] {\scalebox{1.2}{$p_2$}}; % top left leg

  \draw (F) -- ++( 1.697, 1.697) node[right, xshift=4pt] {\scalebox{1.2}{$p_{N-1}$}}; % top right leg
  \draw (F) -- ++( 1.697,-1.697) node[right, xshift=4pt] {\scalebox{1.2}{$p_N$}};     % bottom right leg

  % Top vertical leg labels
  \node at (3,3.0) {\scalebox{1.2}{$p_3$}};
  \node at (6,3.0) {\scalebox{1.2}{$p_4$}};
  \node at (12,3.0) {\scalebox{1.2}{$p_{N-2}$}};

\end{tikzpicture}
\caption{Diagram of the half-ladder residue of the $N$-point scattering amplitude.  This quantity is a function of incoming external momenta $p_1, p_2, \cdots, p_N$, and each internal line corresponds to an on-shell exchanged leg localized to the levels $n_1, n_2, \ldots, n_{N-3}$.  For soft kinematics we set $p_3 = \cdots = p_{N-2} = 0$, in which case all levels  are equal.  For shock wave kinematics we set $p_3 ,\ldots,  p_{N-2} \propto p_2$, in which case these external legs point in the direction of $p_2$ with a democratic superposition over all energies, yielding a uniform sum over all levels.
} \label{fig:ladder}
\end{figure*}
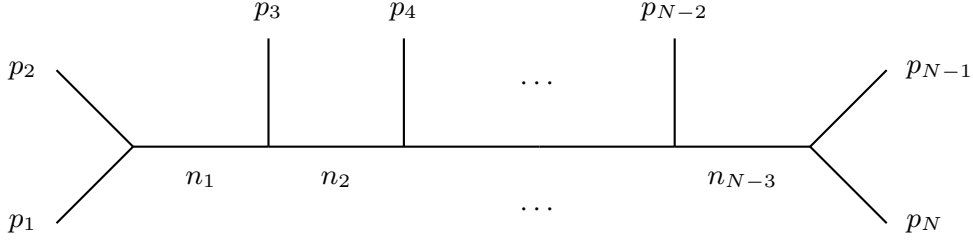

\subsection{Higher-Point Scattering}\label{sec:higher}

Our approach above generalizes to the $N$-point scattering amplitude in the planar limit.   This amplitude depends on $N(N-3)/2$ independent kinematic invariants, which for later convenience we organize according to the intermediate resonances that appear in the half-ladder topology in \Fig{fig:ladder}.  We define the $s$-like invariants,
\eq{
s_i &= -(p_1 + \cdots + p_{i+1})^2 \quad \textrm{for}\quad  1\leq i \leq  N-3 ,
}{}
together with the $t$-like invariants,
\eq{
t_{i,j} & = -(p_i + \cdots + p_j)^2 \quad \textrm{for} \quad 2\leq i,j \leq  N-1  .
}{}
In terms of the variables $X_{i,j} = (p_1 + \cdots + p_{j-1})^2$ in Refs.~\cite{Arkani-Hamed:2023jwn,Arkani-Hamed:2024nzc,BinGeom,CountProblem,Arkani-Hamed:2019mrd,Gluons,Zeros}, the $s$- and $t$-like invariants correspond to $s_i = -X_{1,i+2}$ and $t_{i,j} = -X_{i,j+1}$.
On the maximal cut described by \Fig{fig:ladder}, the corresponding residue is 
\eq{
\hspace{-1.5mm} R_{n_1 n_2 \cdots n_{N{-}3}}\!(t_{i,j}) \,{=} \!  \left[\prod_{i=1}^{N{-}3} \!\!\lim_{s_i{\rightarrow}m_{n_i}^2} \!\!(m_{n_i}^2\!{-}s_i)\!\right] \! A_N(s_i,t_{i,j}).\hspace{-1.5mm}
}{R_def}
Here the $s$-like invariants are taken to be on-shell, while the resulting residue depends on the $t$-like invariants.
Locality forbids an infinite spin tower on any given resonance~\cite{Huang:2022mdb,Caron-Huot:2020cmc,Bern:2021ppb}, which means that the residue is a polynomial in the $t$-like invariants.

\medskip

\noindent {\it   Soft Kinematics.}  Let us consider the kinematic configuration in which all of the ``middle'' legs inserted along the spine of the half-ladder in \Fig{fig:ladder} are taken soft, so
\eq{
p_i= 0\quad \textrm{for} \quad  3\leq i \leq N-2,
}{soft_limit}
with $p_1$, $p_2$, $p_{N-1}$, and $p_N$ hard. In this case, the $N$-point amplitude effectively describes a four-point scattering process in a soft background. This soft limit sets nearly all of the $t$-like invariants to zero,
\eq{
t_{i,j} = 0,
}{tij_0}
except for one of them,
\eq{
t=t_{2,N-1} = -(p_2 +p_{N-1})^2 \neq 0 .
}{t_not0}
Since the middle legs are soft, they do not impart any momentum transfer, so all of the intermediate exchanges must localize to the same level $n$,
\eq{
n_1 = n_2 = \cdots = n_{N-3} = n.
}{}
For notational ease, we define the ``diagonal'' residue,
\eq{
R_{N,n}(t) = R_{\underbrace{\scriptstyle n \,n \, \cdots \,  n}_{\scriptstyle N-3}} (t),
}{}
which at our chosen kinematics is a function only of $t$. 

Taking the $t=0$ forward limit, the diagonal residue can be expressed as an expectation value,
\eq{
R_{N,n}(0) = \langle n | {\cal O}_n^{N-4} | n\rangle,
}{R_Nn}
where $|n\rangle$ is defined in \Eq{n_ket}.  Here we have defined the ``soft operator'' at level $n$,
\eq{
{\cal O}_n &=\sum_{\ell, \ell'}A_{n,n}^{\ell, \ell'}(q_n) |n,\ell\rangle \langle n,\ell'|,
}{O_n}
together with the exchanged momentum,
\eq{
q_n = p_1+p_2,
}{qn_def}
carried by the level $n$ state created by a pair of colliding external massless scalars, so $q_n^2 = -m_n^2$. 
The three-point amplitude $A_{n,n}^{\ell,\ell'}(q_n)$ describes the splitting of an external scalar with zero momentum into a pair of level $n$ states, $|n,\ell\rangle $ and $|n, \ell'\rangle$, both with momentum $q_n$ and taken be in- and out-states, respectively. 
Viewed as the entries of a matrix labeled by $\ell$ and $\ell'$, the three-point amplitude defines a Hermitian matrix,
\eq{
\bar A_{n,n}^{\ell,\ell'}(q_n)=A_{n,n}^{ \ell',\ell}(q_n).
}{}
Physically, this property holds because the soft scalar effectively generates a correction to the  Hermitian propagator matrix for the level $n$ subspace.   The advantage of the representation of the residue in \Eq{R_Nn} is that positivity is manifest.  For instance,  $R_{N,n}(0)\geq 0$ for any even multiplicity $N$.

As before, we can perform a complex momentum shift on $p_1$, $p_2$, $p_{N-1}$, and $p_N$,  where
\eq{
p_1(z)&=p_1+zv, \qquad &p_2(z)&=p_2-zv, \\
p_{N-1}(\bar z) &= -p_2+\bar z\bar v, & p_N(\bar z) &= -p_1 -\bar z\bar v,
}{}
In terms of the states $|n,z\rangle$ in \Eq{nz_ket}, the diagonal residue is then
\eq{
   R_{N,n}(t(z,\bar z))  = \langle n, \bar z|{\cal O}_n^{N-4}| n,z\rangle.
}{R_Nn_tzzbar}
We can also act with derivatives as before to generate new states, yielding an expression for the $\partial^r_t$ derivative of $R_{N,n}(t)$ evaluated at $t=0$,
\eq{
R_{N,n}^{(r)}(0) = \frac{1}{r!}\left( \partial_{\bar z}^r \langle n,\bar z|  \right){\cal O}_n^{N-4}\left(\partial_{ z}^r |n,z\rangle \right) |_{z=\bar z =0},
}{R_Nn_0}
which is itself an expectation value.

We have shown that the diagonal residues of the higher-point amplitude and their derivatives are expectation values of powers of ${\cal O}_n$.  Consequently, these residues satisfy well-known moment inequalities.   For example, all  even moments are obviously positive semidefinite. So too is the variance of ${\cal O}_n$, 
\eq{
\langle {\cal O}_n^2 \rangle - \langle {\cal O}_n\rangle^2 \geq 0,
}{O_variance}
where $ \langle {\cal O }_n^{N-4} \rangle = \langle n,  \bar z|{\cal O}_n^{N-4}| n,z\rangle /  \langle n, \bar z| n,z\rangle$.
More generally, we can define a Hankel matrix of moments,
\eq{
\boldsymbol{{\cal O}}_n =  \left(
\begin{array}{cccc}
1 & \langle {\cal O}_n\rangle & \langle {\cal O}_n^2\rangle & \cdots  \\
\langle {\cal O}_n\rangle & \langle {\cal O}_n^2\rangle & \langle {\cal O}_n^3\rangle  &\\
\langle {\cal O}_n^2\rangle & \langle {\cal O}_n^3\rangle  & \langle {\cal O}_n^4\rangle &  \\
\vdots &  &  &\ddots \\
\end{array}
\right),
}{}
whose entries are $ \langle{\cal O}_n^{i+j}\rangle$, where $i$ and $j$ label rows and columns. 
Since ${\cal O}_n$ is Hermitian, the quantity $\langle {\cal O}_n^{N-4}\rangle$ is real for all $N$. 
For any arbitrary real linear combination of operators, ${\cal O}(u) = \sum_{i=0}^\infty {\cal O}_n^i u_i$, we know that  $\langle {\cal O}(u)^2 \rangle  = \sum_{i,j=0}^\infty u_i \langle  {\cal O}_n^{i+j} \rangle u_j  \geq 0$.  This implies that the matrix $\boldsymbol{{\cal O}}_n$ 
is positive semidefinite,
\eq{
\boldsymbol{{\cal O}}_n \succeq 0,
}{}
so all principal minors of $\boldsymbol{{\cal O}}_n$ are nonnegative.
In terms of the residues, \Eq{O_variance} implies that
\eq{
R_{6,n}(0) R_{4,n}(0) -R_{5,n}(0)^2  \geq 0,
}{simplest_bound}
which is the simplest bound relating data from different multiplicities.  To define the full space of bounds, let us define the residue matrix,
\eq{
\boldsymbol{R}_n(t) =  \left(
\begin{array}{cccc}
R_{4,n}(t) & R_{5,n}(t) & R_{6,n}(t) & \cdots  \\
R_{5,n}(t) & R_{6,n}(t) & R_{7,n}(t)  &\\
R_{6,n}(t) & R_{7,n}(t)  & R_{8,n}(t)  &  \\
\vdots &  &  &\ddots \\
\end{array}
\right).
}{R_n_matrix}
As shown in \Eq{R_Nn_tzzbar}, each entry of this matrix is a moment, so it is positive semidefinite,
\eq{\boldsymbol{R}_n(t)  \succeq 0 \;\;{\rm for}\;\; t\geq 0,
}{}
and equivalently, all of its principal minors are nonnegative. Furthermore, the $t$ derivatives of this matrix at $t=0$ also define  the moments in \Eq{R_Nn_0}, so $\boldsymbol{R}^{(r)}_n(0)$ is itself a positive semidefinite matrix for all $r$.  Putting this all together, we obtain our final constraint,
\eq{
\boldsymbol{R}^{(r)}_n(0)  \succeq 0,
}{R_r_n_pos}
which says that all principal minors of $\boldsymbol{R}^{(r)}_n(0)$ are nonnegative.

Note that for any bound in \Eq{R_r_n_pos}, the $t$ derivatives  act on each residue factor {\it separately}, for example,
\eq{
R_{6,n}^{(r)}(0) R_{4,n}^{(r)}(0) -R_{5,n}^{(r)}(0)^2  \geq 0.
}{Rt_bound}
The above inequality directly implies similar bounds for which $t$ derivatives act {\it outside}  the determinant, 
\eq{
\partial_t^r\bigg(R_{6,n}(t) R_{4,n}(t) -R_{5,n}(t)^2\bigg)\bigg|_{t=0} \geq 0.
}{tR_bound}
This property follows from the fact that if the principal minors of all derivatives of a matrix are nonnegative, then so too are all derivatives of its principal minors.\footnote{The proof of this claim is straightforward.
Assume that a $k$-by-$k$ symmetric matrix $A$ and its derivatives $\partial_t^r A$ are all positive semidefinite.   Without loss of generality we can take the square root of these matrices via  $\partial_t^r A = (B_r)^2$ for some $k$-by-$k$ symmetric matrices $B_r$. From the definition of the determinant, the multinomial theorem, and the product rule, we find that derivatives of $\det A$ can be expressed in a positive semidefinite form,
\begin{equation*}
\begin{aligned}
&\partial_t^r (\det A) \\&= \sum_{r_1 + \cdots +r_k = r}\frac{r!}{r_1!\cdots r_k!}\epsilon_{a_{1}\cdots a_{k}}\epsilon_{b_{1}\cdots b_{k}}(\partial_t^{r_1} A_{a_{1}b_{1}})\cdots (\partial_t^{r_k} A_{a_{k}b_{k}})
\\&= \sum_{r_1 + \cdots +r_k = r}\frac{r!}{r_1!\cdots r_k!} C_{c_1\cdots c_k}C_{c_1\cdots c_k} \geq 0,
\end{aligned}
\end{equation*}
where we defined $C_{a_1\cdots a_k} = \epsilon_{b_1\cdots b_k}(B_{r_1})_{a_1 b_1}\cdots (B_{r_k})_{a_k b_k}$.  
}

Remarkably, we will encounter physical scenarios in which {\it all} of the multipositivity bounds in \Eq{R_r_n_pos} are exactly saturated, which is to say, all principal minors of size greater than one vanish. In these cases, every entry of the residue matrix $\boldsymbol{R}_n(t)$ is equal up to a trivial overall coupling, so $R_{N,n}(t) = \gamma^{N}f(t)$ for some multiplicity-independent function $f(t)$ and constant $\gamma$.  Hence, ${\cal O}_n$ is proportional to the identity matrix, modulo its null space.   From \Eq{O_n}, this reveals a remarkable property of the underlying theory: all three-point couplings between the external massless scalar and any pair of degenerate states $|n,\ell\rangle$ and $|n,\ell'\rangle$ are exactly equal in the soft limit.   This {\it coupling universality} is automatic if there are no degeneracies at level $n$, but otherwise it is quite special. We will return later to the notion of coupling universality in the context of string theory.

\bigskip

\subsection{General Bounds}\label{sec:gen}

The positivity bounds derived above have many generalizations.  
First,  while \Eq{R_Nn} is the expectation value on a {\it pure state} at level $n$, this construction can be extended to an arbitrary positive semidefinite Hermitian density matrix,
\eq{
\rho = \sum_{nn'} \rho_{nn'} |n\rangle \langle n'|,
}{rho}
which describes a statistical mixture of various choices of pure states parameterizing superpositions of levels at the first and last intermediate legs in the half-ladder in \Fig{fig:ladder}.
Requiring our multipositivity bounds to be satisfied for all possible density matrices is mathematically equivalent to imposing them on the set of pure states alone.  However, we will find it convenient here to keep the density matrix completely general. 

Second, we can also generalize the level $n$ soft operator in \Eq{O_n} to the weighted operator,
\eq{
{\cal O}(\sigma) &= \sum_{n,n',\ell, \ell'} \sigma_{nn'} A_{n,n'}^{\ell,\ell'}(q_n,q_{n'}) |n,\ell\rangle \langle n',\ell'|,
}{O_sigma}
where $\sigma_{nn'}$ is any Hermitian matrix and $A_{n,n'}^{\ell,\ell'}(q_n,q_{n'})$ is the three-point amplitude describing the splitting of an external massless scalar into a pair of states, $|n,\ell\rangle $ and $|n', \ell'\rangle$, with momenta $q_n$ and $q_{n'}$ such that $q_n^2 = -m_n^2$ and $q_{n'}^2 = -m_{n'}^2$, respectively. Here the first and second arguments of this three-point amplitude are defined as in- and out-states, respectively.
The three-point amplitude can be viewed as a Hermitian matrix labeled by $n,\ell$ and $n',\ell'$ because
\eq{
\bar A_{n,n'}^{\ell,\ell'}(q_n,q_{n'})=A_{n'n}^{\ell',\ell}(q_{n'},q_n).
}{}
Now recall that the entry-by-entry Hadamard product of a pair of Hermitian matrices generates yet another Hermitian matrix.   Since $\sigma_{nn'}$ and $A_{n,n'}^{\ell,\ell'}(q_n,q_n')$ are both Hermitian, so too is $\sigma_{nn'} A_{n,n'}^{\ell,\ell'}(q_n,q_n')$, which is the definition of the weighted operator ${\cal O}(\sigma)$.

Given the positive semidefinite matrix $\rho$ and the Hermitian matrix  ${\cal O}(\sigma)$, we now define the moment,
\eq{
\widetilde R_{N}(t) ={\rm tr} ( \rho \,  {\cal O}(\sigma)^{N-4}),
}{R_N_gen}
which populates the entries of the moment matrix,
\eq{
\widetilde{\boldsymbol{R}}(t) =  \left(
\begin{array}{cccc}
\widetilde R_{4}(t) & \widetilde R_{5}(t) & \widetilde R_{6}(t) & \cdots  \\
\widetilde R_{5}(t) & \widetilde R_{6}(t) & \widetilde R_{7}(t)  &\\
\widetilde R_{6}(t) & \widetilde R_{7}(t)  & \widetilde R_{8}(t)  &  \\
\vdots &  &  &\ddots \\
\end{array}
\right).
}{R_tilde_matrix}
The definition of $\widetilde R_{N}(t)$ on the right-hand side of \Eq{R_N_gen} implies that the principal minors of $\widetilde{\boldsymbol{R}}(t) $, along with the principal minors of its  $t$ derivatives, are nonnegative for $t\geq 0$, so
\eq{
\boldsymbol{\widetilde R}^{(r)}(0) \succeq 0.
}{gen_bound}
 The very simplest of these bounds imply the nonnegativity of weighted residues at even multiplicity,
 \eq{
\widetilde R_{N}^{(r)}(0) \geq 0 \;\; \textrm{for even } N.
}{}
Furthermore, the inequality  $\det(\boldsymbol{\widetilde{R}}^{(r)}(0))_{456}  \geq 0$ for the $N=4,5,6$ system gives
 \eq{
\widetilde R_6^{(r)}(0) \widetilde R_4^{(r)}(0) - \widetilde R_5^{(r)}(0)^2 \geq 0,
}{}
much like in our earlier analysis.

At this point it would be  tempting to declare victory and assert that we have successfully derived new constraints on the residues of the theory.  However, any such conclusion would be premature: it {\it assumes} that the product of three-point amplitudes defined in \Eq{R_N_gen} is truly equal to the weighted sum of residues, 
\begin{widetext}
\eq{
\widetilde R_{N}(t) = \sum_{n_1=0}^\infty \sum_{n_2=0}^\infty\cdots \sum_{n_{N-3}=0}^\infty    R_{n_1 n_2  \cdots n_{N-3}}(t) \, \rho_{n_{N-3} n_1} \sigma_{n_1 n_2}\sigma_{n_2 n_3}\cdots \sigma_{n_{N-4}n_{N-3}}.
}{R_tilde_sum}
\end{widetext}
At first glance, this condition appears tautologically true.  After all, the maximal cut of any residue is equal to the product of three-point amplitudes.  However, this conclusion implicitly assumes that the values of $q_n$ entering into the operator ${\cal O}(\sigma)$ in \Eq{R_N_gen} correspond to actual {\it physical kinematics}, which is far from automatic.  In particular, ${\cal O}(\sigma)$  depends on $q_n$, but the physical residues depend on the invariants $t_{i,j}$.  Until we draw an explicit connection between the $q_n$ and the $t_{i,j}$, we have not established that the $\widetilde R_{N}(t)$ defined in \Eq{R_N_gen} have any relation whatsoever to physical residues.

Thus we are left with the question: for what choices of $q_n$ do these moments correspond to physical kinematics, and what are the corresponding values of $t_{i,j}$? A comprehensive answer to this question is beyond the scope of this paper, but provides a compelling avenue for future work.  Nevertheless,  there exist simple choices of kinematics for which the above positivity bounds indeed constrain physical residues, which we will now discuss.

\subsubsection{Soft Bounds}\label{sec:soft}

Our aim is to construct an explicit kinematic configuration for which the $N$-point residue is exactly equal to the object defined in \Eq{R_N_gen}.  This choice should correspond to the half-ladder topology of \Fig{fig:ladder}, with a sequence of exchanged states, each at level $n_i$ for $ 1\,\,{\leq}\,\, i\,\,{\leq}\,\, N\,{-}\,3$.  By definition, each exchanged state carries momentum
\eq{
q_{n_i} = p_1 + \cdots + p_{i+1},
}{}
where $q_{n_i}^2 = m_{n_i}^2$.  At the same time, the on-shell condition on the external legs implies that the difference, $q_{n_{i}}  -q_{n_{i-1}}  = p_{i+1}$, must satisfy $p_{i+1}^2 = 0$.

These conditions are trivially satisfied if the difference $q_{n_{i}}  -q_{n_{i-1}}$ vanishes, which corresponds to the soft limit,
\eq{
p_3, \ldots, p_{N-2} =0.
}{}
In this case, $q_{n_i} = p_1 +p_2$  for $1\,\,{\leq}\,\, i\,\,{\leq}\,\, N\,{-}\,3$, coinciding precisely with the kinematics we explored earlier. As before, this configuration requires all levels $n_i$ to be equal and furthermore sets $t_{i,j}=0$ except for $t=t_{2, N-1} \geq 0$.

In fact, the only difference from our earlier analysis in \Sec{sec:higher} is the presence of the weighting matrices $\rho$ and $\sigma$.  Since all intermediate exchanges in the half-ladder are at the same level, we require the weights to be diagonal,
\eq{
\rho_{nn'} = \rho_n \delta_{nn'}\;\;\;{\rm and}\;\;\;\sigma_{nn'} = \sigma_n \delta_{nn'},
}{diag}
for real $\sigma_n$ and real nonnegative $\rho_n$.  In this case, the moment defined in \Eq{R_N_gen} is precisely equal to a weighted sum over physical residues at soft kinematics, so
\eq{
\widetilde R_{N}(t) &\;\;\;\overset{\substack{\textrm{soft}\\ \vspace{.0001cm}}}{=}\;\;\;  \sum_{n=0}^\infty    R_{N,n }(t)\,  \rho_{n} \sigma_n^{N-4}.
}{R_tilde_sum_diag}
Inserting this expression into the moment matrix in \Eq{R_tilde_matrix}, we obtain the positivity bounds in \Eq{gen_bound}.  This construction enforces an infinite web of mixed-multiplicity constraints on weighted sums of residues, where each residue is evaluated with all intermediate exchanges set to the same level.

\subsubsection{Shock Wave Bounds}\label{sec:shock}

It is natural to ask if there is any physical kinematic configuration for which $q_{n_{i}}  -q_{n_{i-1}} \neq 0$.  
One way to achieve such a situation is to set  $q_{n_{i}}  -q_{n_{i-1}}   \propto p_2$, corresponding to
\eq{
p_3, \ldots, p_{N-1} \propto p_2.
}{pprop}
We will refer to this setup as a ``shock wave'' since all legs except $p_1$ and $p_N$ form a jet of energy in the direction of $p_2$.   These collinear kinematics set $t_{i,j}=0$ except for $t=t_{2,N-1}\geq 0 $,  exactly as in the soft limit.   Note that this kinematic configuration is reminiscent of the ``branon'' setup explored in Ref.~\cite{Guerrieri:2024ckc}.

Since $q_{n_{i}}  -q_{n_{i-1}}   \propto p_2$ is nonzero, our choice of kinematics allows for transitions between intermediate levels, which is to say, the $n_i$ are not necessarily the same.  However, to accommodate transitions between different levels, we encounter a new subtlety: the validity of the moment bounds require each insertion of ${\cal O}(\sigma)$ to be the {\it exact same operator}.  If the external scalars are Fock states, then they must carry the same momentum and can thus only accommodate a momentum kick of fixed size.  In this case, a transition between levels is only possible if some mass difference is finely-tuned to this momentum kick.

A way to allow for all transitions among levels is to take a {\it superposition} of Fock states for the external scalars inserted in rungs $3$ through $N-2$ of the half-ladder.
The simplest implementation is a democratic superposition, where each Fock state has equal weight.  This equal superposition happens for a Mellin transform, which integrates uniformly over all energies while always pointing in the direction of $p_2$. Concretely, for the external scalars we choose the state
\eq{
|\psi_i \rangle = \int_{-\infty}^\infty d\omega_i |\omega_i p_2\rangle,
}{ket_Mellin}
where $3\,{\leq}\,\, i \,\,{\leq}\, N\,{-}\,2$.  For these Mellin states, the operator ${\cal O}(\sigma)$ is Hermitian, and \Eq{R_N_gen} describes a residue at genuine physical kinematics.

The discussion above is somewhat involved, but the ultimate conclusion with respect to positivity bounds is simple:
\Eq{R_tilde_sum} genuinely describes a weighted sum over {\it physically realizable residues}, evaluated at all possible combinations of levels $n_i$ for each intermediate state and with $t_{i,j}\,{=}\,0$ except for $t\,{=}\,t_{2,N-1}\,{\geq}\, 0 $.  The corresponding kinematic configuration describes external states that are shock waves comprising Mellin superpositions over energy defined in \Eq{ket_Mellin}, so the intermediate states can transition between levels $n_i$.
While this choice of kinematics superficially matches that of \Eqs{tij_0}{t_not0}, which arise from soft kinematics, in this context it corresponds to Mellin kinematics, which also set all  $t_{i,j}\,{=}\,0$ except for $t\,{=}\,t_{2,N-1}\,{\geq}\, 0 $.
In conclusion, we take the moment matrix $\widetilde{\boldsymbol{R}}(t)$ in \Eq{R_tilde_matrix}, whose elements are the weighted residues $\widetilde R_{N}(t)$ in \Eq{R_tilde_sum}, and for some choice of weights $\rho$ and $\sigma$ extract the infinite set of positivity bounds $\boldsymbol{\widetilde R}^{(r)}_{N}(0) \succeq 0$ described in \Eq{gen_bound}.

In the present analysis, we will sometimes consider the case where the matrix elements of the weights $\rho_{nn'}$ and $\sigma_{nn'}$ are all 0 or 1.  Concretely, suppose $\rho_{nn'} = 1$ for $n,n'\in S_\rho$ and $\sigma_{nn'} = 1$ for $n,n'\in S_{\sigma}$, where $S_\rho \subseteq S_\sigma$ are subsets of the integer levels, with all other entries in $\rho$ and $\sigma$ vanishing. 
For this choice of $\rho_{nn'}$ and $\sigma_{nn'}$, the weighted residue in Eq.~\eqref{R_N_gen}---from which we can derive ``subset'' bounds as  particular special cases of our shock wave kinematics---becomes
\eq{ 
\widetilde R_N(t) &\; \;\overset{\substack{\textrm{subset}\\ \vspace{.0001cm}}}{=}  \sum_{\substack{n_1,n_{N-3}\in S_\rho \\ n_2,n_3,\cdots, n_{N-2} \in S_\sigma}} R_{n_1 n_2 \cdots n_{N-3}}(t),
}{RS}
so the first and last levels are summed over $S_\rho$ while the middle levels are summed over $S_\sigma$.  

Many other choices of weights are also possible.  As we will see shortly, by choosing weights that are given by certain powers of the mass spectrum, we can obtain weighted residues that are exactly equal to the Wilson coefficients of the corresponding low-energy EFT.

\section{Positivity of Wilson Coefficients}

We have derived multipositivity bounds that diagnose a given set of higher-point residues for inconsistency.   Naturally, these residue constraints can be uplifted into constraints on the Wilson coefficients of a general EFT, which we will now describe.

\subsection{Four-Point Scattering}

At low energies, we can expand the tree-level four-point amplitude in powers of $s$ to obtain 
\eq{
A(s,t) = \sum_{k} g_{k}(t) s^{k}.
}{}
Next, we mechanically extract the $t$-dependent Wilson coefficient $g_k(t)$ via the dispersion relation,\footnote{Note that the sum over resonances begins at $n=1$ because the starting contour already encircles the massless $n=0$ pole.}  
\eq{
g_k(t) = \frac{1}{2\pi i}\oint \frac{ds}{s^{k+1}} A(s,t) = \sum_{n=1}^\infty \frac{R_n(t)}{(m^2_n)^{k+1}},
}{}
where the contour of integration is an infinitesimal loop encircling the origin.  In the second equality we have deformed this contour outward, picking up all massive resonances in the $s$ channel.  We have also dropped all boundary contributions, which vanish if the amplitude falls off as
\eq{
\lim_{s \rightarrow \infty} A(s,t) s^{-k} =0.
}{Regge4}
For sufficiently large $k$, this bound on the Regge behavior is relatively weak.  In the tree-level approximation in which resonances are narrow, which we assume throughout, there is a well-defined notion of the spin of each state.  If the spectrum truncates at finite spin, then the dynamics are quantum field theoretic, and \Eq{Regge4} simply requires that $k$ exceed the maximum spin exchanged.  If instead the spectrum is unbounded in spin, then the  dynamics are intrinsically stringy, and we still expect Regge boundedness for some $k$.

From \Eq{R_nr_pos}, we see that
$R_n^{(r)}(0)\geq 0$, which implies that all $t$ derivatives of $g_k(t)$ are nonnegative,
\eq{
g_k^{(r)}(0)\geq 0.
}{g_k_bound}
The above inequality has a long history tracing back to seminal work on analytic dispersion relations \cite{Martin:1965jj,Nicolis:2009qm} and, more recently, the EFT-hedron \cite{Arkani-Hamed:2020blm}.  However, it is worth noting that these past derivations are rather technical in nature, relying on specific positivity properties of Gegenbauer polynomials and propagator poles.  In contrast, our setup exploits a BCFW-like complex momentum shift to build the  wave functions in \Eqs{nz_ket}{nzbar_ket}, which are simultaneously forward while still probing residues in the $t  >0$ regime.

\subsection{Higher-Point Scattering}

We can now expand the $N$-point amplitude in powers of the $s$-like kinematic invariants to obtain 
\eq{
A_N(s_i, t_{i,j}) =\sum_k g_{N,k}(t_{i,j}) (s_1  s_2 \cdots s_{N-3})^k+ \cdots ,
}{}
focusing on terms in which each $s$-like invariant enters with the same power.  For a planar amplitude, each Wilson coefficient $g_{N,k}(t_{i,j})$ is extracted with a multivariable dispersion relation,
\eq{
&g_{N,k}(t_{i,j}) =  \left(\prod_{i=1}^{N-3}\oint \frac{1}{2\pi i}\frac{ds_i}{s_i^{k+1}}\right) A_N(s_i, t_{i,j}) \\
&= \sum_{n_1=1}^\infty \sum_{n_2=1}^\infty\cdots \sum_{n_{N-3}=1}^\infty \frac{R_{n_1 n_2 \cdots n_{N-3}}(t_{i,j})}{(m^2_{n_1}m^2_{n_2}\cdots m^2_{n_{N-3}})^{k+1}},
}{g_Nk_dispersion}
 with a small contour around the origin in each of the $s_i$ variables and dropping all boundary terms.
  In this case, the dispersion relation sums over all poles in the amplitude in $s_1, s_2, \ldots , s_{N-3}$, picking out the half-ladder topology in \Fig{fig:ladder} in the planar limit.
  
Let us comment briefly on the various subtleties of the dispersion relation in \Eq{g_Nk_dispersion}.  First, we emphasize that it employs a {\it sequential} evaluation of contour integrals.    This prescription is well defined because the amplitude is planar and hence only supports poles in the variables $s_1,s_2,\ldots, s_{N-3}$.  Consequently, there is no ambiguity in the ordering of the integrals or the computation of the residues.  A second caveat regards the boundary term from each integral.  Here we have implicitly assumed that we take $k$ sufficiently large that the Regge behavior holds on the amplitude and all of its partial residues.  For example, we can drop the boundary term in the first contour integral evaluation provided
\eq{
\lim_{s_1 \rightarrow \infty} A_N(s_1,\cdots ) s_1^{-k} =0.
}{Regge1}
Similarly, for the second integral evaluation we require
\eq{
\lim_{s_2\rightarrow\infty} \lim_{s_1\rightarrow m_{n_1}^2} (m_{n_1}^2 - s_1)A(s_1,s_2, \cdots) s_2^{-k} =0,
}{Regge2}
and so on for each sequential residue.
For  $k\geq 0$ this property holds for string amplitudes provided  $t_{i,j}<0$, as shown in Ref.~\cite{Arkani-Hamed:2024nzc} in order to prove $N$-point dual resonance.
As noted earlier, the soft limit requires all $t_{i,j}=0$ except $t=t_{2,N-1}\geq 0 $, so we approach this point from below to stay within the region of convergence.   

It is important to note that the sum in \Eq{g_Nk_dispersion} is only convergent if the associated Wilson coefficient $g_{N,k}(t_{i,j})$ is safely finite in the $t_{i,j}\rightarrow 0$ limit.   Conversely, if the EFT amplitude contains terms that are proportional to $(s_1  s_2 \cdots s_{N-3})^k$ with inverse powers of $t_{i,j}$, then the sum will diverge.  To avoid this pathology, we simply choose $k$ sufficiently large that the contributions to $g_{N,k}(t_{i,j})$ from the massless states of highest spin are zero.\footnote{This caveat is familiar from four-point dispersion relations.  For example, the twice-subtracted dispersion relation for graviton scattering extracts the $s^2$ coefficient of the EFT amplitude, which scales as $1/t$ because the amplitude is  $A\sim s^2/t$.  Hence, this dispersion relation equates $1/t$ to an infinite sum of terms that will diverge in the $t=0$ limit.  To ensure convergence, one must instead consider a subtraction of higher degree, which probes EFT coefficients that are finite at $t=0$~\cite{Bellazzini:2015cra}.
Of course, in our setup here, $t$ itself can be taken to be positive, allowing one to sidestep this issue for four-point scattering, but analogous singularities can occur in the higher-point case, which we avoid by choosing $k$ appropriately.
}

Whether \Eqs{Regge1}{Regge2} are valid will of course depend on the theory in question.  
However, in any quantum field theory, $A_N$ is bounded by some power of $s_i$ that depends on $N$ and the highest spin exchanged.  In this case, there always exists some $k$ for which the boundary terms are zero.

Let us set all $t_{i,j}=0$ except for $t = t_{2,N-1}\geq0$,
\eq{
\! g_{N,k}(t) \,{=} \!\! \sum_{n_1=1}^\infty \sum_{n_2=1}^\infty \!\!{\cdots}\!\!\!\! \sum_{n_{N-3}=1}^\infty \! \frac{R_{n_1 n_2 \cdots n_{N-3}}(t)}{(m^2_{n_1} m^2_{n_2}{\cdots}\, m^2_{n_{N-3}})^{k+1}}.\!\!
}{g_Nk_soft}
As in Eq.~\eqref{g_Nk_dispersion}, the residues $R_{n_1 n_2 \cdots n_{N-3}}$ above are defined in \Eq{R_def} by taking a limit of the amplitude on the poles of a physical factorization channel. Next, we observe the literal mathematical equivalence between the Wilson coefficient and a particular weighted residue,
\eq{
 g_{N,k}(t)=\widetilde R_{N}(t),
 }{geqR}
corresponding to a specific choice of weights in Eq.~\eqref{R_tilde_sum}, 
\eq{
\rho_{n,n'} =\sigma_{n,n'}=\frac{1}{m_n^{k+1}  m_{n'}^{k+1}},
}{rho_sigma_Wilson_coeff}
for $n,n'> 0$.  As discussed at length in \Sec{sec:shock}, the weighted residues $\widetilde R_{N}(t)$ satisfy multipositivity bounds because they can arise from a physically realizable kinematic configuration: a process with external states that are in a Mellin superposition over all energies. This democratic sum over states allows for all possible transitions between levels $n_i$ among the external states, which is required in the definition of $\widetilde R_{N}(t)$.  Taking the multipositivity bounds on the weighted residues as input---namely, as a brute mathematical fact about the quantities $\widetilde R_{N}(t)$---the equivalence in Eq.~\eqref{geqR} then directly implies that the same inequalities are satisfied by  $ g_{N,k}(t)$.\footnote{Given that each residue can be viewed as a process involving Mellin states, it is tempting to then conclude that we are somehow computing the Wilson coefficients of the EFT at low energies on the support of these very same Mellin states.  However, this is not necessary.  The multipositivity bounds imply strict constraints on the weighted residues  $\widetilde R_{N}(t)$ as functions of multiplicity $N$ and $t$, which can then be abstracted away from any particular backstory for the precise kinematics of external states.}
In particular, the moment matrices are equivalent,
\eq{
\boldsymbol{g}_{k}(t) =\widetilde{\boldsymbol{R}}(t),
}{g_k_matrix}
where $\widetilde{\boldsymbol{R}}(t)$ is defined in Eq.~\eqref{R_tilde_matrix} with the weights in Eq.~\eqref{rho_sigma_Wilson_coeff}, and the Wilson coefficient matrix is
\eq{
\boldsymbol{g}_k(t) =  \left(
\begin{array}{cccc}
g_{4,k}(t) & g_{5,k}(t) & g_{6,k}(t) & \cdots  \\
g_{5,k}(t) & g_{6,k}(t) & g_{7,k}(t)  &\\
g_{6,k}(t) & g_{7,k}(t)  & g_{8,k}(t)  &  \\
\vdots &  &  &\ddots \\
\end{array}
\right).
}{g_k_matix_def}
Since our shock wave bounds hold for all choices of positive semidefinite $\rho$ and Hermitian $\sigma$, it follows that all principal minors of this $\widetilde{\boldsymbol{R}}(t)$, along with its $t$ derivatives, are nonnegative according to \Eq{gen_bound}, and we learn that the same is true of $\boldsymbol{g}_k(t)$,
\eq{
\boldsymbol{g}^{(r)}_k(0)  \succeq 0.
}{gbound}
This result implies an infinite set of positive semidefinite conditions on the coefficients of the low-energy expansion of the $N$-point scattering amplitude.  The very simplest of these multipositivity bounds imply that the even-point coefficients are nonnegative, so
\eq{
g_{N,k}^{(r)}(0) \geq 0 \;\; \textrm{for even } N.
}{}
The simplest bound involving distinct multiplicities is
\eq{
g_{6,k}^{(r)}(0)g_{4,k}^{(r)}(0) -g_{5,k}^{(r)}(0)^2 \geq 0 ,
}{}
and so on.  These multipositivity bounds answer in the affirmative the question of whether higher-point consistency imposes new constraints on lower-point processes. Concretely, since multipositivity bounds  relate processes at {\it different multiplicities}, knowledge at higher-point directly constrains lower-point and vice versa.

\subsection{General Bounds} \label{sec:gen_bounds_2}

In the previous discussion, we saw how the Wilson coefficients of the EFT can be interpreted as the weighted residue $\widetilde R_N(t)$ in \Eq{R_tilde_sum}, with weights $\rho$ and $\sigma$ given by the powers of the mass spectrum defined in \Eq{rho_sigma_Wilson_coeff}.  However, it is clear that one can derive more general constraints by choosing $\rho$ and $\sigma$ with even more freedom.  For example, consider the weights
\eq{
\!\! \rho_{n,n'}=\sum_{k,k'}\frac{\hat\rho_{k,k'}}{m_n^{k+1}m_{n'}^{k'+1}} ,\;\; \sigma_{n,n'}= \sum_{k,k'}\frac{\hat\sigma_{k,k'}}{m_n^{k+1}m_{n'}^{k'+1}} ,
}{weight_EFT}
for $n,n'> 0$, where $\hat \rho$ is a positive semidefinite Hermitian matrix and $\hat \sigma$ is a Hermitian matrix.
Inserting these matrices into \Eq{R_tilde_sum}, we obtain the weighted residue
\begin{widetext}
\eq{
\widetilde R_{N}(t) &= \sum_{n_1=1}^\infty \sum_{n_2=1}^\infty\cdots \sum_{n_{N-3}=1}^\infty    R_{n_1 n_2  \cdots n_{N-3}}(t) \, 
\sum_{k_1,k_2,\ldots, k_{N-3}} \sum_{k'_1,k'_2,\ldots, k'_{N-3}}
\frac{\hat\rho_{k_{N-3}, k_1'} \hat\sigma_{k_1,k_2'}\hat\sigma_{k_2,k_3'}\cdots \hat\sigma_{k_{N-4},k_{N-3}'}}{m_{n_1}^{k_1+k_1'+2}m_{n_2}^{k_2 + k_2'+2} \cdots m_{n_{N-3}}^{k_{N-3}+ k_{N-3}'+2}
} \\
&= \sum_{k_1,k_2,\ldots, k_{N-3}} \sum_{k'_1,k'_2,\ldots, k'_{N-3}}
\hat\rho_{k_{N-3},k_1'} \hat\sigma_{k_1,k_2'}\hat\sigma_{k_2,k_3'}\cdots \hat\sigma_{k_{N-4},k_{N-3}'}
g_{k''_1 k''_2 \cdots k''_{N-3}}(t) \bigg|_{k_i'' =\frac12 (k_i+k_i')} = \widetilde{g}_N(t).
}{R_tilde_is_Wilson}
\end{widetext}
To derive the second line of \Eq{R_tilde_is_Wilson}, we have exploited the fact that the first line can be recast in terms of Wilson coefficients of the EFT via the dispersion relation
\eq{
 & \hspace{-2mm} g_{k_1 k_2 \cdots k_{N-3}}(t_{i,j}) =  \left(\prod_{i=1}^{N-3}\oint \frac{1}{2\pi i}\frac{ds_i}{s_i^{k_i}}\right)  A_N(s_i, t_{i,j}) \\
& \hspace{-2mm} = \sum_{n_1=1}^\infty \sum_{n_2=1}^\infty\!{\cdots} \!\!\!\! \sum_{n_{N-3}=1}^\infty \! \frac{R_{n_1 n_2 \cdots n_{N-3}}(t_{i,j})}{m^{2(k_1{+}1)}_{n_1}m^{2(k_2{+}1)}_{n_2}{\cdots}\, m^{2(k_{N{-}3}{+}1)}_{n_{N-3}}},\hspace{-5mm}
}{g_dispersion_gen}
dropping all boundary terms as usual.   In terms of the low-energy expansion of the amplitude, these Wilson coefficients enter as
\eq{
&A_N(s_i, t_{i,j})= \\ &\qquad \sum_{k_1,k_2,\ldots, k_{N-3}} \!\!\!\!\!\!\! \!\! g_{k_1 k_2 \cdots k_{N{-}3}}\!(t_{i,j}) s_1^{k_1}  s_2^{k_2} {\cdots} s_{N-3}^{k_{N-3}} {+} \cdots .
}{g_gen_def}
In the second line of \Eq{R_tilde_is_Wilson}, we have defined $\widetilde g_N(t)$ to be the resulting weighted sum of Wilson coefficients, all evaluated at the usual kinematic configuration for which all $t_{ij}=0$ except for $t\,{=}\,t_{2,N-1}\,{\geq}\, 0 $.

Note that in \Eq{weight_EFT}, the entries of $\hat \rho$ and $\hat \sigma$ must be chosen so that all $k_i''$ in \Eq{R_tilde_is_Wilson} are integers wherever the sum has nonzero support; this stipulation ensures that no kinematic invariant enters with fractional powers, so the dispersion relation in \Eq{g_dispersion_gen} is free of branch cuts.
For example, the choice in Eq.~\eqref{rho_sigma_Wilson_coeff} corresponds to setting all entries of $\hat \rho$ and $\hat\sigma$ to zero except for a single diagonal component.

In summary, in \Eq{R_tilde_is_Wilson} we have shown that the weighted Wilson coefficient $\widetilde g_N(t)$ is equal to a weighted residue $\widetilde R_N(t)$.  As a result, we can define the weighted Wilson coefficient matrix,
\eq{
{\boldsymbol{\widetilde{g}}}(t) =  \left(
\begin{array}{cccc}
\widetilde{g}_{4}(t) & \widetilde{g}_{5}(t) & \widetilde{g}_{6}(t) & \cdots  \\
\widetilde{g}_{5}(t) & \widetilde{g}_{6}(t) & \widetilde{g}_{7}(t)  &\\
\widetilde{g}_{6}(t) & \widetilde{g}_{7}(t)  & \widetilde{g}_{8}(t)  &  \\
\vdots &  &  &\ddots \\
\end{array}
\right),
}{g_matrix_weight}
which is then required to be positive semidefinite,
\eq{
\boldsymbol{\widetilde{g}}^{(r)}(0)  \succeq 0.
}{g_matrix_weight_bound}
This freedom offers a huge space of multipositivity bounds on the Wilson coefficients of any EFT. 
For example, choosing $\hat\rho_{kk'} = \delta_{2k}\delta_{2k'}$ and $\sigma_{kk'} = \delta_{2k}\delta_{4k'} + \delta_{4k}\delta_{2k'}$, we find that the determinant inequality $\det(\boldsymbol{\widetilde{g}}^{(r)}(0))_{456}  \geq 0$ for the $N=4,5,6$ system implies
\eq{
&\left(g_{242}^{(r)}(0)  + g_{233}^{(r)}(0)    + g_{323}^{(r)}(0)  + g_{332}^{(r)}(0) \right) g_2^{(r)}(0)\\ & \qquad \geq  \left( g_{23}^{(r)}(0)   + g_{32}^{(r)}(0) \right)^2  .\hspace{-3mm}
}{}
We leave a comprehensive analysis of the full space of constraints to future work.

\section{Examples}

\subsection{Factorizing Amplitudes}

As a check of our bounds, we employ the methods of Ref.~\cite{Arkani-Hamed:2023jwn} to construct a {\it general ansatz} for an arbitrary residue consistent with factorization.    The mechanics of this approach are straightforward: enumerate all possible three-point amplitudes for arbitrary massive spinning particles, and then fuse them together in a half-ladder chain to obtain a ``factorization residue.'' 

To begin, we define the general three-point amplitude of a massless scalar with a $|n,\ell\rangle$ and a $|n',\ell'\rangle$ state, where here $\ell$ denotes spin,
\eq{
\hspace{-1mm} A_{n,n'}^{\ell{,}\ell'}(q_n,q_{n'}\!)\,{=}\!\!\!\!\!\sum_{j=0}^{\min(\ell,\ell')}\!\!\!\!\!\!\lambda_{n,n'}^{\ell,\ell';j}(\!{-}i q_{n'}e)^{\ell{-}\! j} (i  q_{n}e')^{\ell'\!{-}\!j} (e e')^j.\hspace{-1.5mm}
}{lambda}
To compactly represent the on-shell amplitude, we have decomposed the general {\it tensor} polarization for the state $|n,\ell\rangle$ into a tensor product of a  {\it vector} polarization $e$, and likewise for $|n',\ell'\rangle$ and $e'$.  Here the matrix of coupling constants $\lambda_{n,n'}^{\ell,\ell';j} = \lambda_{n',n}^{\ell',\ell;j}$ is symmetric, since swapping $|n,\ell\rangle$ and $|n',\ell'\rangle$ flips the signs of the ingoing and outgoing momenta, along with introducing a factor of $(-1)^{\ell+\ell'}$ associated with swapping the color ordering. The index $j$ labels various kinematic structures that can appear for the same choice of $|n,\ell\rangle$ and $|n',\ell'\rangle$.
Note that our conventions with respect to factors of $i$ are slightly changed from Ref.~\cite{Arkani-Hamed:2023jwn}, but the difference can be absorbed into a redefinition of the couplings.

The three-point amplitudes in \Eq{lambda} are glued together using the all-spin propagators derived in Ref.~\cite{Chandrasekaran:2018qmx} to form the residues of a general theory.
These objects are functions of the kinematic invariants, along with the unknown coupling constants, which we collectively refer to by $\lambda$, in \Eq{lambda}.  Importantly, the factorization residue 
 defines the space of all possible residues, for a given choice of degeneracy in the spectrum, that can arise in a theory that is consistent with the constraints of factorization.  Conversely, by comparing the factorization residue against a residue computed from a putatively consistent amplitude, one can rule out many models using factorization alone.  A limitation of this approach is that in enumerating the couplings $\lambda$, one must explicitly posit the set of spins and degree of degeneracy of the states at each level~\cite{Arkani-Hamed:2023jwn}.

For the present analysis, however, we take the converse position, which is that any choice of real couplings $\lambda$ can be interpreted as the {\it definition} of some theory.   Hence, for all values of $\lambda$, the factorization residues must automatically satisfy our multipositivity bounds.  This observation can be used as a check of our bounds.  Concretely, we have constructed a factorization residue with exchanged particles at levels $n=0,1,2,3$ and spin exchange up to $\ell=3$.  We then computed all generalized multipositivity bounds in \Eq{gen_bound} for the $N=4,5,6$ system on several thousand randomly sampled choices of couplings $\lambda$, spectrum $m_n^2$, and spacetime dimension $4\leq D \leq 26$.   For this analysis we also chose random values for the density matrix $\rho$ and the weight matrix $\sigma$ in the bounds.  At every sample point, we found that the residue matrix in  \Eq{gen_bound}  for the $N=4,5,6$ system is positive semidefinite for all $t\geq 0$.   We arrived at the same  conclusion studying the $N=4,5,6,7,8$ system at levels $n=0,1,2$ and spin exchange up to $\ell=2$.

For the residues with all exchanges at $n=1$ and $\ell\leq 1$, we can actually show positive semidefiniteness analytically.  The corresponding residues at soft kinematics are
\eq{
R_{N,1}(t) &= (\lambda_{0,1}^{0,0;0})^2 (\lambda_{1,1}^{0,0;0})^{N-4} \\&\;\;\; +\frac{1}{4}(\lambda_{0,1}^{0,1;0})^2 (\lambda_{1,1}^{1,1;1})^{N-4}(m_1^2 + 2t).
}{R_fact_1}
Inserting this formula into the residue matrix in \Eq{R_n_matrix}, we see immediately that all principal minors are positive semidefinite.  By inspection, we see that \eq
{R_{N,1}^{(r)}(0)\geq 0  \;\; \textrm{for even } N.
}{}
 Meanwhile, the next principal minor corresponds to
\eq{
&\det ( \boldsymbol{R}_1(0))_{456} = R_{6,1}(0)R_{4,1}(0){-R}_{5,1}(0)^2 \\& \quad=\tfrac{1}{4}m_1^2 (\lambda_{0,1}^{0,0;0}\lambda_{0,1}^{0,1;0})^2 (\lambda_{1,1}^{0,0;0}-\lambda_{1,1}^{1,1;1})^2 \geq 0,
}
{}
which is the manifestly positive semidefinite determinant of the $n\,\,{=}\,\,1$ residue matrix of the $N\,\,{=}\,\,4,5,6$ system.
Here we see explicitly how saturation of the bounds implies coupling universality in the soft limit when $\lambda_{1,1}^{0,0;0}=\lambda_{1,1}^{1,1;1}$.  The $t$ derivative versions of the above bounds are also satisfied, so for example
\eq{
\det ( \boldsymbol{R}_1^{(1)}(0))_{456} \,{=}\,R_{6,1}^{(1)}(0)R_{4,1}^{(1)}(0)\,{-}\,R_{5,1}^{(1)}(0)^2 \,{=}\, 0 ,
}
{}
and so on.  For the $n=1$ residues defined in \Eq{R_fact_1}, the multipositivity bounds beyond $N=4,5,6$ do not introduce new constraints.

\subsection{String Amplitudes}

The $N$-point string amplitude\footnote{More precisely, by ``string amplitude'' we refer to the planar Koba-Nielsen factor~\cite{Koba:1969rw} describing the open string at tree level, with the Regge intercept chosen to vanish. This so-called $Z$-theory model has been proposed as a self-consistent amplitude~\cite{Ztheory} and describes the bosonic string amplitude at shifted kinematics for which the scalars are massless.} and all of its residues can be efficiently computed using the methods introduced recently in Ref.~\cite{Arkani-Hamed:2024nzc}. Let us briefly review the amplitude and residue computation.  The string amplitude has the compact worldsheet representation,
\eq{
A_{N}^{\rm str} = \int \frac{d^n z}{{\rm SL}(2,\mathbb{R})} \frac{\prod_{i<j}u_{i,j}^{X_{i,j}}}{z_{1,2}\cdots z_{N,1}} ,
}{stringamp}
where we have defined the $u$-variables~\cite{BardakciRuegg,ChanTsou,Koba:1969rw,Gross:1969db},
\eq{
u_{i,j} = \frac{z_{i-1,j}z_{i,j-1}}{z_{i,j}z_{i-1,j-1}},
}{}
in terms of the moduli  parameters $z_i$ of the disc and their pairwise differences $z_{i,j} = z_i - z_j$.

Note that the $u$-variables are not independent, but rather satisfy nonlinear $u$-equations, 
\eq{u_{i,j} + \prod_{(i',j')\cap (i,j)}u_{i',j'} = 1,
}{}
where the product is over chords $(i',j')$ intersecting the chord $(i,j)$.
To automatically satisfy the $u$-equations, we can introduce the positive $y$-variables as in Refs.~\cite{BinGeom,CountProblem,Arkani-Hamed:2019mrd,Gluons,Arkani-Hamed:2023jwn,Zeros,Splits,Arkani-Hamed:2024nzc}, in terms of which the $u_{i,j}$ for $1\leq i\leq N-2$ and $i+2\leq j\leq N$ satisfy 
\eq{
u_{i,j} = \frac{{\cal F}_{i-1,j}{\cal F}_{i,j-1}}{{\cal F}_{i,j}{\cal F}_{i-1,j-1}},
}{yvar}
where we have defined ${\cal F}_{0,j}=\prod_{k=3}^j y_{1,k}$, ${\cal F}_{i,N} = 1$, and otherwise ${\cal F}_{j,k}= 1+\sum_{l=j+2}^k\prod_{m=l}^k y_{1,m}$.

After transforming from $u$-variables to $y$-variables, the integral representation of the amplitude becomes~\cite{Arkani-Hamed:2024nzc}
\eq{
A_{N}^{\rm str} &= \int_0^\infty \prod_{i=3}^{N-1}\frac{dy_{1,i}}{y_{1,i}}y_{1,i}^{X_{1,i}} \prod_{j=1}^{N-3}\prod_{k=j+2}^{N-1} {\cal F}_{j,k}^{-c_{j,k}},
}{stringamp2}
where $c_{j,k} = X_{j,k}+X_{j+1,k+1}-X_{j+1,k}-X_{j,k+1}$.
Here \Eq{stringamp2} evaluates the full amplitude, even though we have made a choice of coordinates for the $y$-variables associated with the half-ladder topology so that we can straightforwardly compute the half-ladder residues.

Using this representation of the amplitude, we can extract any residue $R_{n_1 n_2 \cdots n_{N-3}}$ by simply computing the residue at the origin of each $y$-variable, with the kinematic invariants associated with each pole set to the corresponding mass value.
Taking soft kinematics, which correspond to setting all $t_{i,j}=0$ except for $t = t_{2,N-1}\geq 0$, we obtain  
\eq{
R_{N,n}(t) & = \frac{\Gamma(t+n+1)}{\Gamma(n+1)\Gamma(t+1)}.
}{universal_string_residue}
Note that this expression is remarkably {\it independent} of the multiplicity $N$.   Consequently, the residue matrix $\boldsymbol{R}^{(r)}_n(0)$ has the same value in all of its entries, and all of its principal minors of size greater than one are zero.  In turn, 
the bound $\boldsymbol{R}^{(r)}_n(0)  \succeq 0$ is exactly saturated.  We thus learn that the  string exhibits the coupling universality described in Sec.~\ref{sec:higher}. This accords with the results of Ref.~\cite{Arkani-Hamed:2023jwn}, where the coupling of a massless scalar to a pair of scalars or vectors at level $n=1$ was found to be the identity matrix in the soft limit.

As shown in Ref.~\cite{Arkani-Hamed:2024nzc}, the residues of the open string in the forward limit are $R_{n_1 n_2 \cdots n_{N-3}}(0) = 1$, and consequently Eq.~\eqref{g_Nk_soft} implies that the Wilson coefficients have a striking relation to the Riemann zeta function,
\eq{
g_{N,k}(0) = \zeta(k+1)^{N-3},
}{gzeta}
which clearly agrees with our bounds in Eq.~\eqref{gbound} with a trivial Hankel matrix built from powers of $\zeta(k+1)$.
We have confirmed Eq.~\eqref{gzeta} numerically for $N=5$ and $k=1,2$ using the convergent ${}_6 F_5$ form of the five-point string amplitude found in Ref.~\cite{Arkani-Hamed:2024nzc}.

As an example verifying our more general multipositivity bounds, consider the weighted residue in Eq.~\eqref{R_tilde_sum}. Again using that $R_{n_1 n_2 \cdots n_{N-3}}(0) = 1$, we find that the weighted residue takes a simple form,
\eq{
\widetilde{R}_N(0)={\rm tr}(\rho\sigma^{N-4}).
}{R_N_weight_string}
As expected, this automatically satisfies our multipositivity bounds in Eq.~\eqref{gen_bound} because it is literally the expectation value of powers of a matrix $\sigma$ on a  state $\rho$.
Following the discussion of Sec.~\ref{sec:gen_bounds_2}, we choose $\rho$ and $\sigma$ as in \Eq{weight_EFT} in order to relate this weighted residue to Wilson coefficients of the EFT.
For the string, the EFT coefficients defined in Eqs.~\eqref{g_dispersion_gen} and \eqref{g_gen_def} take the form of a product of Riemann zeta functions evaluated at different arguments,
\eq{
g_{k_{1}k_{2}\cdots k_{N-3}}(0) =\prod_{i=1}^{N-3}\zeta(k_{i}+1).
}{}
The real symmetric matrix $\zeta_{kk'}=\zeta(\tfrac{k+k'}{2}+1)$, defined for $k,k'\geq 1$, is positive definite,\footnote{The positive semidefiniteness of $\zeta_{kk'}$ can be derived by taking the product with an arbitrary real vector $u$, 
\begin{equation*}
u_k \zeta_{kk'}u_{k'}=\sum_{k=1}^\infty \sum_{k'=1}^\infty \sum_{n=1}^{\infty}\frac{u_k u_{k'}}{n^{\frac{k+k'}{2}+1}}=w^2\geq 0,
\end{equation*}
where the vector $w$ has entries $w_n = \sum_{k = 1}^\infty u_k/n^{(k+1)/2}$.} so it has a unique symmetric positive definite square root matrix defined by $  \zeta_{k k'}= Z_{k k''} Z_{k'' k'}$.
The weighted residue in Eq.~\eqref{R_tilde_is_Wilson} with weights chosen as in Eq.~\eqref{weight_EFT} then takes the form
\eq{
\widetilde{g}_{N}(0)={\rm tr}((Z\hat\rho Z)(Z\hat\sigma Z)^{N-4}).
}{g_N_weight_string}
Since $\hat \rho$ is positive semidefinite and both $\hat\rho$ and $\hat\sigma$ are Hermitian, these respective properties are inherited upon conjugation by $Z$.
As a result, the weighted Wilson coefficient matrix in Eq.~\eqref{g_matrix_weight} given for the string in Eq.~\eqref{g_N_weight_string} is positive semidefinite---by the same argument as for $\widetilde{R}_N(0)$ in Eq.~\eqref{R_N_weight_string}---as our multipositivity bounds in Eq.~\eqref{g_matrix_weight_bound} demand.
We leave more general computation of the string Wilson coefficients to future work, but they are guaranteed to satisfy our multipositivity bounds.

\subsection{Bespoke Amplitudes}

The bespoke construction of Ref.~\cite{Cheung:2023uwn} generates dual resonant amplitudes with a ``designer spectrum,''
\eq{
m^2_n \,{=}\, \frac{P(n)}{Q(n)} \,{=}\, \frac{n^h {+} P_1 n^{h-1} {+}\cdots {+} P_{h-1} n {+} P_h}{n^{h-1} {+} Q_2 n^{h-2} {+}\cdots {+} Q_{h-1} n {+} Q_h},
}{spectrum_bespoke}
where $P(n)$ and $Q(n)$ are polynomials of degree $h$ and $h\,{-}\,1$, respectively, described by the constant parameters $P_{i}$ and $Q_{i}$.  These choices of degree ensure that the asymptotic spectrum is linear, as expected for consistency~\cite{Caron-Huot:2016icg,Bhardwaj:2024klc}. 
Throughout, we will assume that the spectrum satisfies
\eq{
m^2_0 =0 \quad \textrm{and} \quad m^{2\prime}_n >0,
}{mass_bound_bespoke}
where the prime denotes a derivative of the mass spectrum with respect to the level $n$, i.e., $m_n^{2\prime} = dm_n^2/dn$.
The first condition ensures that the lowest-lying state is massless, as we have assumed for the external scalars.  The second condition is required for the four-point residues at level $n$ to be nonzero.  In particular, the level~$n$ residue is proportional to the factor $m^{2\prime}_n$ relative to the string, so this quantity cannot be negative.   On the other hand, if $m^{2\prime}_n$ is zero, then the level $n$ modes do not even couple to the external scalar, which is a trivial case we will not consider.  Given the spectrum in \Eq{spectrum_bespoke}, the conditions in \Eq{mass_bound_bespoke} imply that
\eq{
P_h =0 \quad \textrm{and} \quad P_{h-1} \neq 0,
}{p_bound_bespoke}
which we henceforth assume for our analysis.

Next, let us define the planar kinematic invariants $X_I$ for $I=1,\ldots,N(N-3)/2$, which are exactly  the same as the objects $-s_i$ and $-t_{i,j}$ defined in this paper.  In terms of these variables, the $N$-point bespoke amplitudes are
\eq{A_N(X_I) = d_N \sum_{\alpha_I} A_{N}^{\rm str}(-\nu_{\alpha_I}(-X_I)),
}{bespokedef} 
where $d_N =h^{-N(N-3)/2}(m^{2\prime}_0/h)^{N/2} $ is a constant and 
 $A_{N}^{\rm str}$ is the $N$-point string amplitude in Eq.~\eqref{stringamp}.  Here the functions $\nu_\alpha(\mu)$, for $\alpha=0,\ldots,h-1$, are the $h$ roots of the polynomial $P(\nu)-\mu Q(\nu)=0$.
To compute the bespoke amplitude we start with the string amplitude, replace $X_I\rightarrow -\nu_{\alpha_I}(-X_I)$, and then sum over all $h$ roots $\nu$ for each of the $N(N-3)/2$ kinematic invariants.
All branch cuts cancel, and the resulting object is meromorphic, with simple poles and polynomial residues at the locations dictated by the spectrum given in Eq.~\eqref{spectrum_bespoke}, and for certain choices of parameters is dual resonant.
By construction, the $N$-point bespoke amplitude factorizes automatically on the poles at level $n=0$.

The residues at level $n=1$ have a simple formula,
\eq{
R_{N,1}(t) &= c_{N,1} \left(\frac{t}{h}+\frac{1-(-P_1/h)^{N-2}}{1+(P_1/h)} \right) ,
}{}
where we have defined \eq{
 c_{N,n} = \left(\frac{m^{2\prime}_n }{h} \right)^{N-3}\left(\frac{m^{2\prime}_0}{h} \right)^{N/2},
 }{}
which is a positive factor that cancels out of all bounds.
Here the first and second factors arise from Jacobians appearing on the $N-3$ internal legs and $N$ external legs of the half-ladder topology.
 
 The simplest multipositivity bound is computed from the determinant of the $n=1$ residue matrix for the $N=4,5,6$ system,
 \eq{
\det ( \boldsymbol{R}_1(0))_{456} &
\propto -\frac{P_1^2}{h^2} \geq 0,
 }{}
dropping positive multiplicative constants.  The above constraint can only be satisfied if we require that
 \eq{
 P_1=0.
 }{}  
 Given the conditions in \Eq{p_bound_bespoke}, we see that this constraint is inconsistent for the $h=2$ bespoke models, so these theories violate unitarity.   This conclusion accords with the factorization analysis of Ref.~\cite{Arkani-Hamed:2023jwn}, but we have now ruled out the $h=2$ theories, for massless external states, independent of any assumptions about the degeneracy.
 
 The residues at level $n=2$ are more complicated, but can be written in closed form, where we have set $P_1=0$ as per the previous constraint, 
 \eq{
&R_{N,2} (t)= c_{N,2}\Bigg\{\frac{1}{2h}t^2 + \frac{1}{h}\left(\frac{3}{2} + Q_2\right)t + 1
\\&\;\;\;  +\sum_{k=0}^{N-4}\left(-\frac{P_2}{h}\right)^{k+1}{}_{2}F_1 \left[\begin{array}{c}-k,k-N+4 \\ 1\end{array};4\right] \Bigg\}.}{}
The simple $t$ derivative bound $R_{N,2}^{(1)}(0) \geq0$ implies that
\eq{
Q_2 \geq -3/2.
}{}
Meanwhile, the determinant of the $n=2$ residue matrix for the $N=4,5,6,7,8$ system at $t=0$ is
\eq{
\det ( \boldsymbol{R}_2(0))_{45678} &  \propto - \frac{P_2^5}{h^5}  \geq 0,
 }{}
 which implies the inequality
 \eq{
 P_2\leq 0,
 }{}  
 which holds for all values of $h$.

Finally, as a demonstration of our reweighted bounds from Sec.~\ref{sec:gen}, we consider the prescient choice for the weights $\rho_{nn'}$ and $\sigma_{nn'}$ in Eqs.~\eqref{rho} and \eqref{O_sigma} given by 
\eq{
\rho_{nn'} &= \left(\frac{\delta_{2n}\delta_{2n'}}{m_2^{2\prime}} + \frac{\delta_{2n}\delta_{3n'}}{\sqrt{m_2^{2\prime}m_3^{2\prime}}}\right)+\left(2\leftrightarrow 3\right)\\
\sigma_{nn'} &= \left(\delta_{2n}\delta_{2n'}\sqrt{\frac{m_3^{2\prime}}{m_2^{2\prime}}}  - \delta_{2n}\delta_{3n'} \right) + \left( 2\leftrightarrow 3\right) ,
}{}
in which case the weighted residue yields
\eq{
\widetilde R_{6}^{(2)}(0) \propto (1+Q_2) \geq 0.
}{}
Thus, we conclude that
\eq{
Q_2 \geq -1,
}{}
again for all values of $h$.  There are likely many more constraints on the parameter space of bespoke amplitudes at general $h$, whose fate we leave to future work.

\subsection{Deformed Worldsheet Amplitudes}

As a final example, we turn to a broad category of deformations of the $N$-point string amplitude obtained by multiplying the integrand of Eq.~\eqref{stringamp} by a form factor $P_N(u)$, which is a function of the $u_{i,j}$ variables.
Such deformations include all possible ``satellite constructions'' whereby a new amplitude is built from weighted sums of the usual string amplitude evaluated at integer-shifted kinematics.
Some natural choices for $P_N(u)$ were considered and ruled out via an explicit factorization construction in Ref.~\cite{Arkani-Hamed:2023jwn}.

An infinite class of $P_N(u)$ theories was introduced long ago by Gross in Ref.~\cite{Gross:1969db}. 
The simplest version of this construction introduces an exponential form factor,
\eq{\! P_N(u)\,{=}\,\exp\!\left[\,\sum_{k=1}^{N} u_{k,k+2}(1\,{-}\,u_{k,k+2})\tilde f_4(u_{k,k+2})\right]\! ,}{DavidP}
for some unknown function $\tilde f_4(x)$,  where the indices on the $u_{i,j}$ are cyclically defined modulo $N$.
String theory itself corresponds to the case where all $P_N(u)=1$, in which case $\tilde f_4(x)=0$.
More complicated exponential $P_N(u)$ theories are also constructed in Ref.~\cite{Gross:1969db} from functions $\tilde f_{N'}$ for $N'>4$, which describe further deformations of $N$-point string scattering invisible for $N<N'$.  We emphasize here that the theory in Eq.~\eqref{DavidP} describes a deformation of the string amplitude at any $N$-point. 
Interestingly, the amplitudes in Ref.~\cite{Gross:1969db} are {\it guaranteed} to factorize at $N$-point, albeit with couplings that may be imaginary and with dramatically enhanced degeneracy.
Hence, these deformed worldsheet models are ideal test subjects for our bounds, which apply for any choice of degeneracies.

Beyond the $n=0$ diagonal residues, which are simply equal to one as in string theory, the $N$-point residues $R_{n_1 n_2\cdots n_{N-3}}$ depend on $\tilde f_4(x)$ evaluated at $x=0,1$, along with its $x$ derivatives up to order $\max(n_i)-1$.
We can then use multipositivity bounds to constrain the form of $\tilde f_4(x)$. For example, we find that the $n=1$ residue matrix for the $N=4,5,6$ system gives the bound
\eq{
\det ( \boldsymbol{R}_1(0))_{456} &= R_{6,1}(0)R_{4,1}(0) - R_{5,1}(0)^2 \\&= -\tilde f_4(1)^2 ,
}{}
which implies that
\eq{
\tilde f_4(1)=0.
}{}
Meanwhile, the $n=1$ residue matrix for the $N=4,6,8$ system produces the constraint
\eq{
\det ( \boldsymbol{R}_1(0))_{468} &= R_{8,1}(0)R_{4,1}(0) - R_{6,1}(0)^2 \\
&= -\tilde f_4(0)^4 (1+2\tilde f_4(0))^4 ,
}{f401}
while the general residue in  \Eq{RS} for level subsets $S_\rho = \{0\}$ and $S_\sigma = \{0,1\}$ in the $N=4,6,8$ system gives
\eq{
\det ( \widetilde{\boldsymbol{R}}_1(0))_{468} &= \widetilde R_8(0)\widetilde R_{4}(0) - \widetilde R_{6}(0)^2 \\& = 2(1{+}\tilde f_4(0))(2 {+} 6 \tilde f_4(0) {+} 3 \tilde f_4(0)^2).
}{f402}
All together, these conditions imply that
\eq{
\tilde f_4(0)=0.
}{}
In conclusion, our multipositivity bounds prove that all $N$-point residues of the deformed worldsheet theory in Eq.~\eqref{DavidP} coincide precisely with those of string theory up to and including all level $n=1$ exchanges.

More generally, our multipositivity bounds on all subsets of levels up through $n=2$ imply that
\eq{
\tilde f_4'(0) = \tilde f_4'(1) = 0,
}{}
so all $N$-point residues of the deformed theory match the string for all exchanges at level $n=2$ and below.
This conclusion holds as a consequence of unitarity, regardless of the enhanced degeneracy of this model, and offers further evidence for the rigidity of string theory.
It is tempting to conjecture that going to yet higher levels would imply that $\tilde f_4$ vanishes identically, but we leave the investigation of this possibility to future work.

\newcommand{\SixPtForkLadder}{
\begin{tikzpicture}[baseline=-0.1cm, line width=0.8pt, scale=0.25]

  % Define horizontal points
  \coordinate (A) at (0,0);
  \coordinate (B) at (3,0);
  \coordinate (C) at (6,0);
  \coordinate (D) at (9,0);

  % Draw horizontal backbone with labels below
  \draw (A) -- node[below, yshift=-6pt] {} (B);
  \draw (B) -- node[below, yshift=-6pt] {} (C);
  \draw (C) -- node[below, yshift=-6pt] {} (D);

  % Vertical upward legs
  \draw (B) -- ++(0,2.0);  % leg 3 vertical
  \draw (C) -- ++(0,2.0);  % leg 4 vertical

  % Forks at tops of vertical legs
  \draw (3,2.0) -- ++(-1.2,1.2);  % left fork from 3
  \draw (3,2.0) -- ++(1.2,1.2);   % right fork from 3

  \draw (6,2.0) -- ++(-1.2,1.2);  % left fork from 4
  \draw (6,2.0) -- ++(1.2,1.2);   % right fork from 4

  % Diagonal legs on the left
  \draw (A) -- ++(-2,2);   % leg 1
  \draw (A) -- ++(-2,-2);  % leg 2

  % Diagonal legs on the right
  \draw (D) -- ++(2,2);    % leg 5
  \draw (D) -- ++(2,-2);   % leg 6

\end{tikzpicture}
}

\newcommand{\FivePtForkLadder}{
\begin{tikzpicture}[baseline=-0.1cm, line width=0.8pt, scale=0.25]

  % Define horizontal points
  \coordinate (A) at (0,0);
  \coordinate (B) at (3,0);
  \coordinate (C) at (6,0);

  % Draw horizontal spine with labels below
  \draw (A) -- node[below, yshift=-6pt] {} (B);
  \draw (B) -- node[below, yshift=-6pt] {} (C);

  % Vertical upward leg
  \draw (B) -- ++(0,2.0);  % leg 3 vertical

  % Fork at top of vertical leg
  \draw (3,2.0) -- ++(-1.2,1.2);  % left fork from 3
  \draw (3,2.0) -- ++(1.2,1.2);   % right fork from 3

  % Diagonal legs on the left
  \draw (A) -- ++(-2,2);   % leg 1
  \draw (A) -- ++(-2,-2);  % leg 2

  % Diagonal legs on the right
  \draw (C) -- ++(2,2);    % leg 4
  \draw (C) -- ++(2,-2);   % leg 5

\end{tikzpicture}
}

\newcommand{\FourPtForkLadder}{
\begin{tikzpicture}[baseline=-0.1cm, line width=0.8pt, scale=0.25]

  % Define horizontal points
  \coordinate (A) at (0,0);
  \coordinate (B) at (3,0);

  % Draw horizontal spine with label below
  \draw (A) -- node[below, yshift=-6pt] {} (B);

  % Diagonal legs on the left
  \draw (A) -- ++(-2,2);   % leg 1
  \draw (A) -- ++(-2,-2);  % leg 2

  % Diagonal legs on the right
  \draw (B) -- ++(2,2);    % leg 3
  \draw (B) -- ++(2,-2);   % leg 4

\end{tikzpicture}
}

\bigskip

\section{Discussion}

In this paper, we have derived an infinite tower of multipositivity bounds that nonlinearly constrain the space of $N$-point amplitudes and residues under modest assumptions.   The key insight of our approach is to reinterpret the half-ladder residues depicted in \Fig{fig:ladder} as {\it moments}.  Each moment is the expectation value of various powers of an operator corresponding to each external scalar inserted in the middle of the half-ladder, evaluated on the states created by the pairs of external scalars at the ends of the half-ladder.   
The positive semidefinite inequalities satisfied by the moments of any Hermitian matrix then imply a web of nonlinear constraints on all $N$-point residues. Using dispersion relations, these residue inequalities can be recast into bounds on the $N$-point Wilson coefficients of a general planar scalar EFT.     

In the simplest version of our bounds, each operator insertion is associated with a soft external scalar, in which case the corresponding $N$-point residues are effectively evaluated at four-point kinematics.  For more complicated kinematics, which we dubbed shock wave configurations, even stronger bounds can be derived.
These more general bounds involve multiple mass levels, as well as arbitrary weights on the internal operator insertions and initial and final pairs of scalars.  All together, our constraints form an intricate network of consistency  conditions on any residue.
Armed with the totality of multipositivity bounds, we then analyzed---and in some cases ruled out---a variety of deformed string amplitudes.

Our analysis opens up several avenues for future work.  First and foremost is the task of mapping out the full space of multipositivity bounds for general external kinematics.  We have focused on kinematics corresponding to external scalars for legs $3$ through $N-2$ that are either soft or in a Mellin-transformed superposition of energies.  These results still leave a vast space of viable kinematics unexplored, which could easily yield stronger bounds on $N$-point residues and EFT coefficients.  A related question concerns other topologies.  In this paper, we have focused exclusively on the case of half-ladder residues.  However, the basic logic---that insertions of particles can be viewed as operators that satisfy moment inequalities---is more general.  For instance, we should expect bounds on residues and Wilson coefficients of the schematic form
\begin{widetext}
\eq{
\left( \SixPtForkLadder\right) \left(\FourPtForkLadder\right) -\left(\FivePtForkLadder\right)^2 \geq 0,
}{}
\end{widetext}
where each fork along the spines of these diagrams is interpreted as another operator insertion.

A second open question concerns the span of all possible positivity bounds on the Wilson coefficients ${g}_{N,k}^{(r)}(0)$.  These couplings correspond to the kinematic dependence of EFT amplitudes at a given multiplicity $N$,  power $k$ in $s$-like invariants, and power $r$ in $t$-like invariants.  Past works on positivity such as the EFT-hedron have focused solely on four-point scattering, incorporating moment bounds that mix different values of $k$ and $r$.  The analysis presented in this paper has centered on moment bounds that mix different values of $N$, at fixed $k$ and $r$.  Hence, an obvious target of opportunity is to synthesize all positivity bounds in $N$, $k$, and $r$ within some unified framework.

Last but not least is the question of planar versus nonplanar theories.  Our multipositivity bounds on residues apply to the specific half-ladder topology---which appears in any theory with cubic interactions---so here planarity is not required.  In contrast, our bounds on EFT coefficients are derived by extracting the half-ladder topology of massive exchanges from a dispersion relation, and this construction relies crucially on planarity to avoid nonplanar poles.  It is therefore natural to ask whether a nonplanar construction of our multipositivity bounds on EFT coefficients also exists.   

Notably, even for the planar case, there are further constraints that we have yet to fully explore.  For example, consider the four-point dispersion relation,
\eq{
A(s,t)&= g' (s t)^k {+}\cdots,\\
{\rm where} \;\; g'&= \oint \frac{ds}{s^{k+1}}  \frac{dt}{t^{k+1}} A(s,t) ,
}{}
for some positive integer $k$.
In string theory---or more generally, any theory describing an infinite spin tower---the Regge behavior at large $s$ depends on $t$, and thus the contours in those variables cannot be deformed to infinity, since there is boundary term.  However, for any tree-level quantum field theory describing a finite tower of spin exchanges, there is some $k$ for which this boundary term is zero.  In this case, the dispersion relation localizes to terms with a pole in the $s$ and $t$ channels simultaneously.   For a planar theory, no such pole exists, and we conclude that $g'=0$.  
The converse of this logic is remarkable: if one {\it observes} a nonzero value of $g'$ at large $k$, then we can immediately deduce that either the ultraviolet completion has no upper bound on the spin of the exchanged states---so it is stringy or loop-level---or that it violates Lorentz invariance, unitarity, or analyticity. Said another way, the observation of nonzero $g'$ would preclude any tree-level quantum field theory.

\bigskip
\bigskip
\bigskip
\bigskip
\bigskip

\noindent {\it Acknowledgments:} 
We thank Nima Arkani-Hamed, Justin Berman, Henriette Elvang, David Gross, Aaron Hillman, and Sasha Zhiboedov for conversations.
We additionally thank the referee for thoughtful feedback.
C.C.~is supported by the Department of Energy (Grant No.~DE-SC0011632) and by the Walter Burke Institute for Theoretical Physics.  
G.N.R.~is supported by the James Arthur Postdoctoral Fellowship at New York University.

\newpage

\bibliographystyle{utphys-modified}
\bibliography{multipositivity}

\vspace{-10mm}

\end{document}